\documentclass{article}
\usepackage{amssymb}
\usepackage{fullpage}
\usepackage{amsmath}
\usepackage{amsfonts}
\usepackage{graphicx}
\usepackage{float}

\input{tcilatex}

\begin{document}

\title{The Stability of an Isotropic Cosmological Singularity in
Higher-Order Gravity}
\author{Jonathan Middleton and John D. Barrow \\
DAMTP, Centre for Mathematical Sciences, \\
Cambridge University, \\
Wilberforce Rd., Cambridge CB3 0WA, UK}

\maketitle

\begin{abstract}
We study the stability of the isotropic vacuum Friedmann universe in gravity
theories with higher-order curvature terms of the form $(R_{ab}R^{ab})^{n}$
added to the Einstein-Hilbert Lagrangian of general relativity on approach
to an initial cosmological singularity. Earlier, we had shown that, when $%
n=1$, a special isotropic vacuum solution exists which behaves like the
radiation-dominated Friedmann universe and is stable to anisotropic and
small inhomogeneous perturbations of scalar, vector and tensor type. This is
completely different to the situation that holds in general relativity,
where an isotropic initial cosmological singularity is unstable in vacuum
and under a wide range of non-vacuum conditions. We show that when $n\neq 1$,
although a special isotropic vacuum solution found by Clifton and Barrow
always exists, it is no longer stable when the initial singularity is
approached. We find the particular stability conditions under the influence
of tensor, vector, and scalar perturbations for general $n$ for both solution branches. On approach to the initial singularity, the isotropic vacuum solution with scale factor $a(t)=t^{P_{-}/3}$ is found to be stable to tensor perturbations for $0.5<n< 1.1309$ and stable to vector perturbations for $0.861425 < n \leq 1$, but is unstable as $t \rightarrow 0$ otherwise. The solution with scale factor $a(t)=t^{P_{+}/3}$ is not relevant to the case of an initial singularity for $n>1$ and is unstable as $t \rightarrow 0$ for all $n$ for each type of perturbation.
\end{abstract}

\section{\protect\bigskip Introduction}
The study of the very early universe leads us to investigate what happens to
our assumptions about the truth of Einstein's general theory of relativity
when the curvature of space and the density of matter and radiation
approach the fundamental Planck values defined by the constants of Nature,
$G,c$ and $h$. The most natural extensions to explore
as generalisations of general relativity are the higher-order theories of
gravity that arise when the Einstein-Hilbert Lagrangian is extended by adding powers of the scalar curvature or
the square of the Ricci tensor. As the Planck epoch is reached, or passed, on approach to
a cosmological singularity, these higher-order terms are expected to
dominate the behaviour of simple cosmological models. Any evaluation of what
are likely initial conditions during the pre-inflationary era of a
cosmological model should therefore be based on a full understanding of the
general behaviour of cosmological models in the presence of higher-order
gravity terms.

Contributions to the Lagrangian from powers of the scalar curvature, $R^n$, are conformally equivalent to the presence of a
self-interacting scalar field and are understood \cite{barcot}. In an earlier paper \cite{mid}, we considered the effect on cosmological
singularities of adding the quadratic Ricci invariant $R_{ab}R^{ab}$ to the
Einstein-Hilbert action of general relativity. The purely quadratic
Lagrangian gravity theories that contain this invariant, but not the
Einstein-Hilbert ($R$) term, possess an isotropic vacuum cosmological
solution, in which the expansion scale factor, $a(t)$, behaves as in the
flat Friedmann radiation-dominated universe of general relativity, with $%
a(t)=t^{1/2}$ \cite{special, bher1}. In the case of zero spatial curvature\footnote{%
The Friedmann radiation solutions are also exact solutions of the pure $%
R_{ab}R^{ab}$ theory in the cases of non-zero spatial curvature \cite{bher1}.}%
, this vacuum solution of the pure $R_{ab}R^{ab}$ theory therefore has the
exact metric:
\begin{eqnarray}
ds^2=-dt^2+t(dx^2+dy^2+dz^2) . \label{iso} \end{eqnarray}
Thus, we see that the higher-order Ricci stresses induce a behaviour that
mimics the effect of an isotropic black-body radiation stress, even though
no physical stress of this sort is present. Earlier studies of anisotropic,
spatially homogeneous universes of Bianchi types I, II \cite{bher1} and IX
\cite{cotIX} showed that this special isotropic solution is stable against homogeneous anisotropic distortions as $%
t\rightarrow 0$. This surprising situation is completely different to that
encountered in general relativity (GR), when the $R_{ab}R^{ab}$ term is
absent from the action. In GR, the expansion and 3-curvature anisotropies
dominate the vacuum dynamics as $t\rightarrow 0$ so as to produce
anisotropic \cite{misner, bkl}, and even chaotic \cite{misnercc}, dynamics. For all perfect
fluids with pressure, $p$, and density, $\rho $, satisfying $-\rho /3<p<\rho
$, the isotropic solution is unstable as $t\rightarrow 0$ and hence such isotropic solutions are special in GR \cite{stab}. This instability
does not occur when the $R_{ab}R^{ab}$ term is present. On approach
to the cosmological singularity, the higher-order curvature terms render the
isotropic solution stable. This has all sorts of consequences for physical
cosmology. For example, it ensures that a pre-inflationary state will likely
be isotropic and it removes the need for the introduction of an extra
physical principle, like the minimisation of a `gravitational entropy' \cite%
{pen2, bher2}, in order to enforce a special isotropic initial state.
However, it does suggest that a stable state of isotropic contraction will be
produced on approach to any future singularity in a closed universe and that
may be an awkward conclusion for any theory of a gravitational entropy
governed by its own gravitational `Second Law'.

The addition of quadratic Ricci terms can also create unusual evolutionary
behaviour, not seen in general relativity. Barrow and Hervik found exact
solutions which display anisotropic inflation \cite{bher1, bher2}. These
solutions do not have a general-relativistic limit and are intrinsically
non-linear with respect to the space-time curvature.

In our first paper \cite{mid}, we extended the study of the effects of an $%
R_{ab}R^{ab}$ addition to the Einstein-Hilbert action to the situation of
anisotropic and inhomogeneous cosmologies. Specifically, we investigated the
behaviour of small scalar, vector, and tensor perturbations to the metric %
(\ref{iso}) as $t\rightarrow 0$. We found that there were \textit{no} growing
metric perturbation modes of scalar, vector, or tensor sorts as $%
t\rightarrow 0$. Thus, a small perturbation of the isotropic cosmological
solution forms part of the general solution of the gravitational field
equations when the $R_{ab}R^{ab}$ term is present: it is an open property of
the initial data space of the quadratic theory.

These results immediately suggest that we should investigate whether or not the
stability of isotropic singularities is maintained to higher order when we
introduce additions to the Einstein-Hilbert action of the form $(R_{ab}R^{ab})^{n}$. We expect the situation for $n\neq 1$ to be more
complicated because there will no longer be a simple Gauss-Bonnet invariant
underlying the field equations. This question of the stability of the $n\neq
1$ theories is the subject of this paper. In the absence of the
Einstein-Hilbert term there is a counterpart to the simple isotropic vacuum
solution of equation (\ref{iso}) in the case of general $n$, which
was found by Clifton and Barrow \cite{Clifton2}. This reduces to the
solution (\ref{iso}) as $n\rightarrow 1$. It is the stability of this
isotropic power-law solution for general $n$ that we shall investigate.

In section \ref{eqns} we give the field equations for the gravity theory with an $%
R+A(R_{ab}R^{ab})^{n}$ Lagrangian and give the exact isotropic vacuum
solutions. These solutions have two branches. We identify the physically
interesting one that describes an expanding universe and show that as $%
n\rightarrow \infty $ the exact vacuum solution approaches that of a
dust-filled general relativity solution with $a(t)=t^{2/3}$.

In section \ref{perts}, we present the formalism for studying small tensor, vector and
scalar perturbations of this special vacuum solution in order to determine
the conditions on $n$ for which it is stable as $t\rightarrow 0$ and the
initial singularity is approached. In sections \ref{tensor}, \ref{vector}, and \ref{scalar} these stability
analyses are carried out for tensor, vector, and scalar perturbation modes,
respectively. The results are summarised and discussed in section \ref{summary}. A
collection of useful quantities is derived in the appendices.

\section{Field Equations} \label{eqns}
Consider a higher-order gravity theory with action
\begin{equation*}
S=\int d^{4}x\sqrt{-g}\left[ \frac{1}{\chi }(R+A(R_{ab}R^{ab})^{n})+L_{m}%
\right] ,
\end{equation*}%
where $\chi, A$ and $n$ are constants. The field equations are obtained using the
general formula from Clifton and Barrow \cite{Clifton2} which expresses the
higher-order contributions as an additional effective stress tensor:

\begin{equation}
G_{b}^{a}+AP_{b}^{a}=\frac{\chi }{2}T_{b}^{a}\ ,
\end{equation}%
where
\begin{eqnarray*}
P_{b}^{a} & \equiv &-\frac{1}{2}Y^{n}g_{b}^{a}+nR_{b}^{a}\Box
(Y^{n-1})+nY^{n-1}\Box R_{b}^{a}+2ng^{cd}(Y^{n-1})_{,c}R_{b;d}^{a} \\
&&+ng_{b}^{a}\left( Y^{n-1}\!_{;cd}R^{cd}+2(Y^{n-1})_{,c}R^{cd}\!_{;d}+\frac{%
1}{2}Y^{n-1}\Box R\right) \\
&&-n\biggl((Y^{n-1})_{;b}\!\,^{c}R_{c}^{a}+(Y^{n-1})_{;}\,\!^{a}\,%
\!_{c}R_{b}^{c}+(Y^{n-1})_{;b}R^{ca}\!_{;c}+g^{ad}(Y^{n-1})_{;d}R_{b;c}^{c}
\\
&&+(Y^{n-1})_{,c}R^{ca}\!_{;b}+(Y^{n-1})_{,c}g^{ad}R^{c}_{b;d}\,%
\!^{a}+Y^{n-1}(g^{ad}R_{;db}+2R^{a}\,\!_{cdb}R^{cd})\biggr),
\end{eqnarray*}%
with $Y=R^{ab}R_{ab},$ and $G_{ab} \equiv R_{ab}-\frac{1}{2}Rg_{ab}$ is the usual
Einstein tensor.

We consider perturbations about a spatially flat, homogeneous and isotropic
FRW spacetime with metric
\begin{equation}
ds^{2}=-dt^{2}+a^{2}(t)(dx^{2}+dy^{2}+dz^{2}),
\end{equation}%
with aforementioned scale factor $a(t)$ and associated Hubble expansion rate
$H\equiv \frac{\dot{a}}{a}$.

In the limit where the Ricci term dominates, $A\rightarrow \infty $, which
we expect to be appropriate in the neighbourhood of the cosmological
singularity where $a\rightarrow 0$, provided $n>\frac{1}{2}$, the vacuum field equations reduce to $%
P_{b}^{a}=0$. To background order, we have:
\begin{eqnarray}
P_{0}^{0} &=&-Y^{n-2}\biggl(\frac{1}{2}Y^{2}+6nY(2H\ddot{H}-2\dot{H}%
^{2}+3H^{2}\dot{H}-3H^{4}) +6n(n-1)\dot{Y}(2H\dot{H}+3H^{3})\biggr), \\
P_{\alpha }^{0} &=&0\;=\;P_{0}^{\alpha }, \\
P_{\beta }^{\alpha } &=& -Y^{n-3}\delta _{\beta }^{\alpha }\biggl\{\frac{1}{2}%
Y^{3}+nY^{2}\left( 4\dddot{H}+24H\ddot{H}+12\dot{H}^{2}+18H^{2}\dot{H}%
-18H^{4}\right)  \notag \\
&&+n(n-1)Y\dot{Y}\left( 8\ddot{H}+36H\dot{H}+12H^{3}\right)+2n(n-1)\left( 2\dot{H}+3H^{2}\right) \left( (n-2)\dot{Y}^{2}+Y\ddot{Y}%
\right)  \biggr\}.
\end{eqnarray}
Substituting for $Y$ and $\dot{Y}$ in terms of $H,\dot{H},\ldots $ gives
\begin{eqnarray*}
P_{0}^{0} &=&-72(12\dot{H}^{2}+36H^{2}\dot{H}+36H^{4})^{n-2}\biggl\{\dot{H}%
^{4}+6H^{2}\dot{H}^{3}+15H^{4}\dot{H}^{2}+18H^{6}\dot{H}+9H^{8} \\
&&+n(-2H\dot{H}^{2}\ddot{H}-6H^{3}\dot{H}\ddot{H}-3H^{5}\ddot{H}-2\dot{H}%
^{4}-15H^{2}\dot{H}^{3}-42H^{4}\dot{H}^{2}-36H^{6}\dot{H}-9H^{8}) \\
&&+n^{2}(4H\dot{H}^{2}\ddot{H}+12H^{3}\dot{H}\ddot{H}+9H^{5}\ddot{H}+12H^{2}%
\dot{H}^{3}+42H^{4}\dot{H}^{2}+36H^{6}\dot{H})\biggr\}.
\end{eqnarray*}%
Hence, the Friedmann-like equation for this theory in vacuum is
\begin{eqnarray}
0 &=&H^{5}\ddot{H}(9n^{2}-3n)+H^{3}\dot{H}\ddot{H}(12n^{2}-6n)+H\dot{H}^{2}%
\ddot{H}(4n^{2}-2n)+\dot{H}^{4}(1-2n)  \notag \\
&&+H^{2}\dot{H}^{3}(6-15n+12n^{2})+H^{4}\dot{H}%
^{2}(15-42n+42n^{2}) +H^{6}\dot{H}(18-36n+36n^{2})+H^{8}(9-9n).
\end{eqnarray}%

For power-law scale factors, $a=t^{k}$, and general values of $n\neq 1,$this
implies \begin{equation*}
k=0,k=\frac{1}{2}\pm \frac{i}{6}\sqrt{3},\text{ or }k=\frac{P}{3},
\end{equation*}%
where the possible values of $P$ are given by the two roots of a quadratic:
\begin{equation}
P=P_{\pm }=\frac{3(1-3n+4n^{2})\pm \sqrt{3(-1+10n-5n^{2}-40n^{3}+48n^{4})}}{%
2(1-n)}\;.
\end{equation}

\begin{figure}[h]
\centering \includegraphics[scale=0.5]{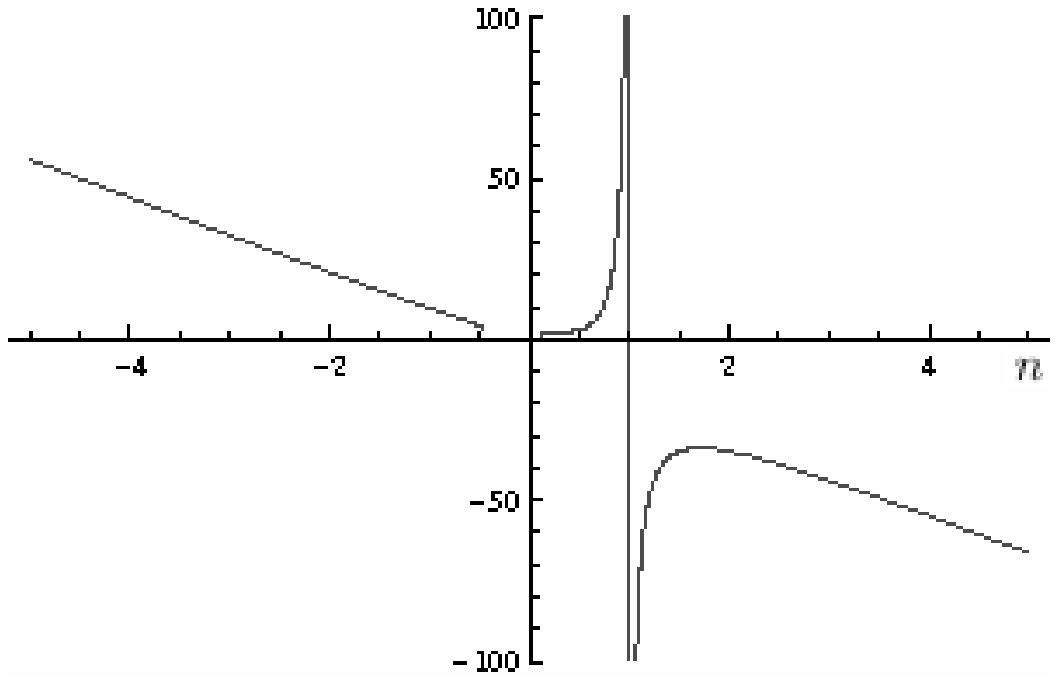}
\caption{The variation of $P_{+}$ with $n.$}
\label{fig:Pp}
\end{figure}

\begin{figure}[h]
\centering \includegraphics[scale=0.5]{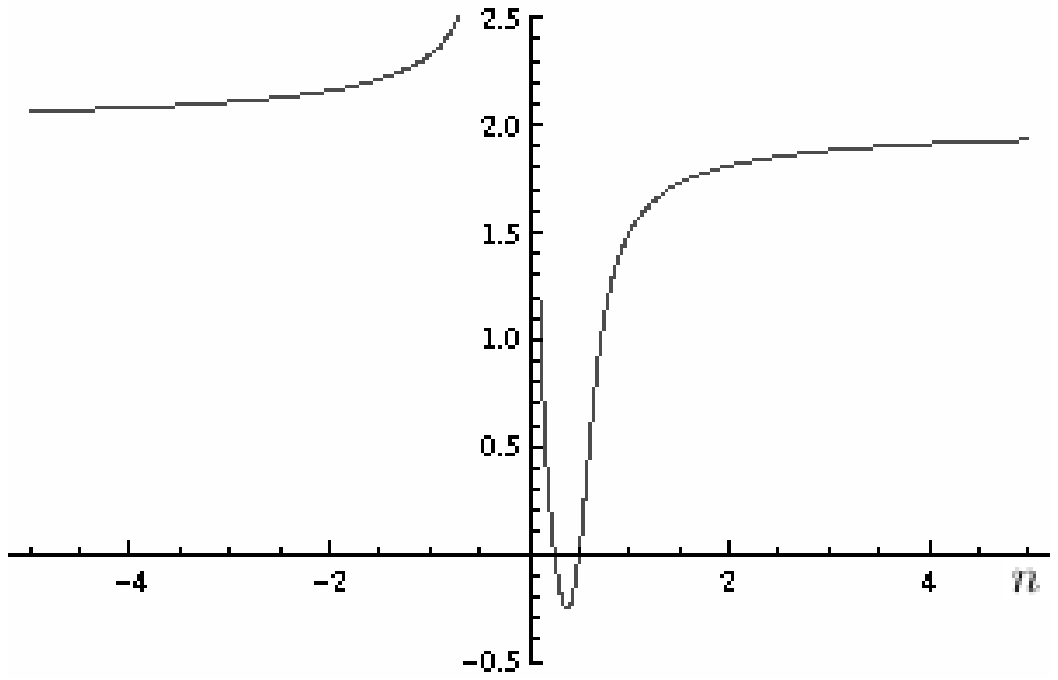}
\caption{The variation of $P_{-}$ with $n.$}
\label{fig:Pm}
\end{figure}

In the limit $n\rightarrow 1$, $P_{-}$ $\rightarrow \frac{3}{2}$, and we
obtain the special $a=t^{\frac{1}{2}}$ vacuum solution of the quadratic ($%
n=1 $) case studied in ref. \cite{mid}. Note also that $P_{-}$ rapidly
asymptotes towards $2$ as $n\rightarrow \infty $,
\begin{eqnarray}
P_{-}\rightarrow 2-\frac{1}{3n}-\frac{1}{18n^{2}}-\frac{13}{216n^{3}}+ O(n^{-4}),  \label{asym}
\end{eqnarray}%
and the vacuum solution rapidly approaches the behaviour of the GR dust
solution with $a=t^{2/3}$, see equation (\ref{asym}). $P$ (or its real part) is
greater than $3$ only for the range $-\frac{1}{2}<n<-0.390388$. For the
choices $k=0,k=\frac{1}{2}\pm \frac{i}{6}\sqrt{3}$, we must have $n>0$,
since $12\dot{H}^{2}+36H^{2}\dot{H}+36H^{4}$ also vanishes. The physically
interesting cases relevant to an initial singularity are those with $k>0$,
i.e. solutions which are expanding to the future. Finally, we note that an
exponential scale factor with $H=constant$ is possible iff $n=1$.

For comparison, in a perfect fluid-filled universe with equation of state $%
p=w\mu $, there is a flat FRW exact solution of the $(R_{ab}R^{ab})^{n\text{
\ }}$theory where the scale factor is given by
\begin{equation}
a(t)=t^{\frac{4n}{3(w+1)}} .
\end{equation}

\section{Inhomogeneous Perturbations} \label{perts}
We will now develop the formalism for studying small perturbations of the spatially flat isotropic FRW solutions of the $(R_{ab}R^{ab})^{n\text{
\ }}$theory, which generalises the formalism
developed by Noh and Hwang for the quadratic ($n=1$) theory \cite{Noh:gw, Noh:vort, Noh:scalar}.
We are interested in the stability of the spatially flat isotropic
background FRW solution
\begin{eqnarray}
a(t)=t^{\frac{P}{3}},   \label{nsoln}
\end{eqnarray}%
where
 \begin{equation}
P=P_{\pm }=\frac{3(1-3n+4n^{2})\pm \sqrt{3(-1+10n-5n^{2}-40n^{3}+48n^{4})}}{%
2(1-n)}\;.
\end{equation}
The general perturbed metric may be written as
 \begin{eqnarray}
ds^{2} &=& -a^{2}(1+2\alpha )d\eta ^{2}-a^{2}\tilde{B}_{\alpha }d\eta dx^{\alpha
} +a^{2}(\delta _{\alpha \beta }+\tilde{C}_{\alpha \beta })dx^{\alpha
}dx^{\beta },
\end{eqnarray}
where $\eta $ is a conformal time coordinate that is related to the comoving
proper time, $t$, by $dt=ad\eta $. We can decompose the perturbation
variables into their scalar, vector and tensor parts in the standard way, as
in \cite{mid}, by writing
\begin{eqnarray*}
\tilde{B}_{\alpha } &=&2\beta _{,\alpha }+2B_{\alpha }, \\
\tilde{C}_{\alpha \beta } &=&2\phi \delta _{\alpha \beta }+2\gamma _{,\alpha
\beta }+2C_{(\alpha ,\beta )}+2C_{\alpha \beta }.
\end{eqnarray*}

There are four scalar perturbation variables, $\alpha ,\beta ,\phi $ and $%
\gamma $, two vector variables, $B_{\alpha }$ and $C_{\alpha },$ and one
tensor, $C_{\alpha \beta }$. The quantities $B_{\alpha }$ and $C_{\alpha }$
are divergence-free, i.e. $B^{\alpha }\!_{,\alpha }\equiv 0\equiv C^{\alpha
}\!_{,\alpha }$, and $C_{\alpha \beta }$ is transverse and trace-free. These
three types of perturbation evolve independently of each other at linear
order. We will determine the equations which describe their time evolution
and then solve each of them to determine whether the metric perturbations to
the special solution are stable as $t\rightarrow 0$. In the $n=1$ case the
problem, the equations, and their solutions will reduce to those of \cite%
{mid}. In this way we establish the ranges of $n$ values for which the
special isotropic vacuum solution is a stable initial condition for the
higher-order theory.

\section{Tensor (gravitational-wave) perturbations} \label{tensor}

The expansion of the metric around the spatially flat Friedmann solution now
takes the form

\begin{equation*}
ds^{2}=-dt^{2}+a^{2}(\delta _{\alpha \beta }+2C_{\alpha \beta })dx^{\alpha
}dx^{\beta }.
\end{equation*}%
The tensor $C_{\alpha \beta }$ is trace-free and transverse, i.e.
\begin{equation}
C_{\alpha }^{\alpha }=0=C_{\beta ,\alpha }^{\alpha }  \label{con}
\end{equation}%
and $C=C(\mathbf{x},t)$ .

The $n=1$ case was solved exactly in \cite{mid} for perturbations about $a(t)=t^{%
1/2}$. Here, we want to perturb an isotropic background
solution which has $a(t)=t^{P/3}$ with
\begin{equation}
P=P_{\pm }=\frac{3(1-3n+4n^{2})\pm \sqrt{3(48n^{4}-40n^{3}-5n^{2}+10n-1)}}{%
2(1-n)}.
\end{equation}%
The important quantities to linear order in the perturbation are given in Appendix \ref{app:t}. In the limit where the higher-order terms dominate, the perturbed field
equation is

\begin{eqnarray}
\delta P_{\beta }^{\alpha } &=&-nY^{n-1}\biggl(\ddddot{C}_{\beta }^{\alpha }+6H\dddot{C}_{\beta }^{\alpha
}+3H^{2}\ddot{C}_{\beta }^{\alpha }-(3\ddot{H}+21H\dot{H}+18H^{3})\dot{C}%
_{\beta }^{\alpha } -2\frac{\Delta }{a^{2}}\ddot{C}_{\beta }^{\alpha }-2H\frac{\Delta }{a^{2}}%
\dot{C}_{\beta }^{\alpha }\notag
\\
&& +\left( 4\dot{H}+8H^{2}+\frac{\Delta }{a^{2}}%
\right) \frac{\Delta }{a^{2}}C_{\beta }^{\alpha }\biggr) -n(Y^{n-1})^{\textbf{.}}\left( 2\dddot{C}_{\beta }^{\alpha }+9H\ddot{C}_{\beta
}^{\alpha }+3H^{2}\dot{C}_{\beta }^{\alpha }-2\frac{\Delta }{a^{2}}\dot{C}%
_{\beta }^{\alpha }+H\frac{\Delta }{a^{2}}C_{\beta }^{\alpha }\right)  \notag
\\
&& -n(Y^{n-1})^{\textbf{..}} \left( \ddot{C}_{\beta }^{\alpha }+3H\dot{C}_{\beta
}^{\alpha }-\frac{\Delta }{a^{2}}C_{\beta }^{\alpha }\right)  .
\end{eqnarray}

\subsection{Large Scales}

In the long-wavelength limit, on super-horizon scales, we can neglect terms
involving $\Delta C$, $\Delta ^{2}C$ and $\Delta \dot{C}$.

For $a(t)=t^{P/3}$, we have $H=\frac{P}{3t}$ and $Y(t)\propto t^{-4}$%
, so the equation for the perturbations becomes

\begin{eqnarray}
0 \, = \, \delta P_{\beta }^{\alpha } &=&-nY^{n-1}\biggl\{\ddddot{C}_{\beta
}^{\alpha }+(-8n+8+2P)\frac{\dddot{C}_{\beta }^{\alpha }}{t}  +\left( 16n^{2}-28n+12+12P-12Pn+\frac{P^{2}}{3}\right) \frac{\ddot{C}%
_{\beta }^{\alpha }}{t^{2}} \notag \\ && +\left( 16Pn^{2}-28Pn+10P+\frac{11P^{2}}{3}%
- \frac{4P^{2}n}{3}-\frac{2P^{3}}{3}\right) \frac{\dot{C}_{\beta }^{\alpha }%
}{t^{3}}\biggr\}
\end{eqnarray}%
and so
\begin{eqnarray}
C &\propto &t^{\lambda }  \notag \\
0 &=&\lambda \biggl(\lambda ^{3}+(-8n+2+2P)\lambda ^{2}+\left(
-4n-1+16n^{2}+6P(1-2n)+\frac{P^{2}}{3}\right) \lambda  \notag \\
&&-2+12n-16n^{2}+P(2-16n+16n^{2})+P^{2}\left( \frac{10}{3}-\frac{4n}{3}%
\right) -\frac{2}{3}P^{3}\biggr) .
\end{eqnarray}%
The four roots of this are
\begin{equation}
\lambda =0,\;\lambda _{1},\;\lambda _{\pm } ,
\end{equation}%
where
\begin{eqnarray} \label{lambda}
\lambda _{\pm } &=&\frac{1}{2}(\lambda _{1}\pm \sqrt{\lambda _{2}}), \\
\lambda _{1} & \equiv &-1-P+4n \label{lambda1} \\
\lambda _{2} & \equiv &\frac{11}{3}P^{2}-14P+8Pn+9-24n+16n^{2} \label{lambda2} \\
&=&\frac{464n^{4}-392n^{3}+9n^{2}+64n-13\pm (36n^{2}-11n-3)\sqrt{%
3(48n^{4}-40n^{3}-5n^{2}+10n-1)}}{2(1-n)^{2}} \notag .
\end{eqnarray}%
$\lambda _{1}$ and $\lambda _{2} $ are real whenever $P$ is real, i.e. $%
n\not\in (-0.47942,0.110873)$ \footnote{%
The values here and in the tables that follow are approximate numerical
roots of the appropriate polynomials.}. For $\lambda _{\pm }$ to be real we
need $\lambda _{2}\geq 0$.

We are interested in the signs of the possible values of $\lambda $ in order
to determine the behaviour of gravitational-wave perturbations of the
isotropic solution as $t\rightarrow 0$. If any $\Re {(\lambda _{i})}<0$, the
solution is unstable as $t\rightarrow 0$. Otherwise, we need to look at the
stability problem to second order, due to the presence of the zero
eigenvalue. For reference, we recall that for the $n=1$ theory, studied
earlier \cite{mid}, we had
\begin{equation*}
C(\mathbf{x},t)\propto \alpha +\beta t^{1/2}+\gamma t+\delta t^{3/2}
\end{equation*}%
and there were no diverging metric perturbation modes as $t\rightarrow 0$.
Let us now analyse the situation in the more complicated $n\neq 1$ case:

\subsection{Solutions with $P=P_{+}$}

First consider the case $P=P_{+}$, so that
\begin{equation*}
P = \frac{3(1-3n+4n^{2})+\sqrt{3(-1+10n-5n^{2}-40n^{3}+48n^{4})}}{2(1-n)} , \end{equation*}
for which the stability is decided by the quantities \begin{eqnarray*}
\lambda _{1} &=&\frac{-20n^{2}+19n-5-\sqrt{3(-1+10n-5n^{2}-40n^{3}+48n^{4})}%
}{2(1-n)}, \\
\lambda _{2} &=&\frac{464n^{4}-392n^{3}+9n^{2}+64n-13+(36n^{2}-11n-3)\sqrt{%
3(-1+10n-5n^{2}-40n^{3}+48n^{4})}}{2(1-n)^{2}}.
\end{eqnarray*}

\begin{figure}[h]
\centering \includegraphics[scale=0.5]{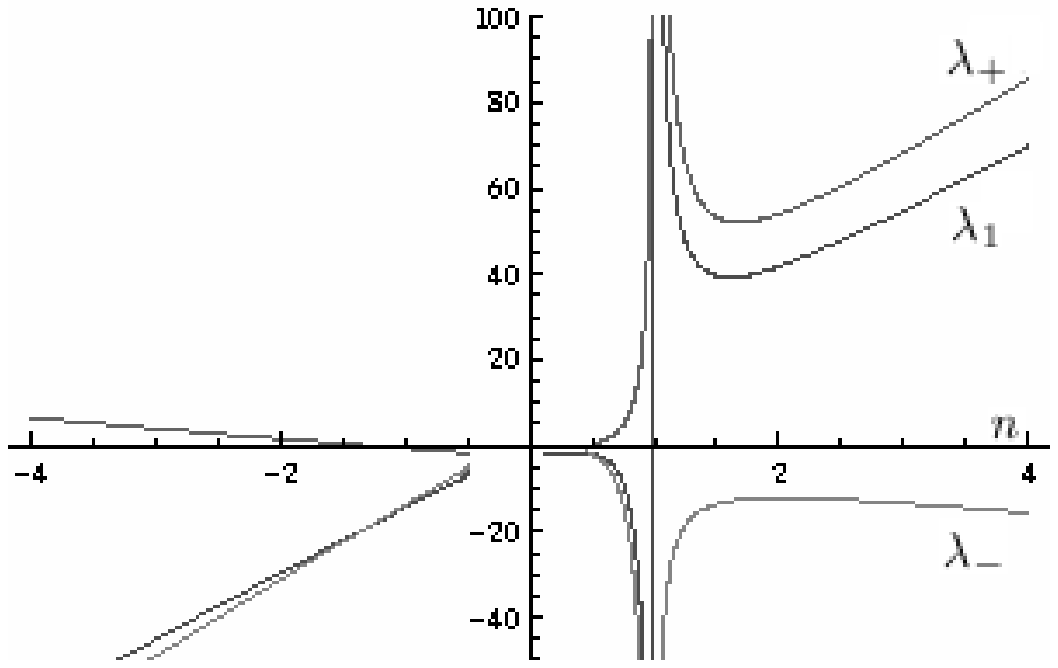}
\caption{Power-law exponents, $\lambda_{1}, \lambda_{\pm}$, versus $n$ for tensor perturbations with $%
P=P_{+}$.}
\label{fig:gplus}
\end{figure}

 \begin{tabular}{|c||c|c|c|cc||r|}
\hline
\multicolumn{7}{|c|}{Values of $\lambda _{i}$ for different values of $n\neq
1$, taking $P=P_{+}$} \\
& $P_{+}$ & $\lambda _{1}$ & $\lambda _{2}$ & $\lambda _{+}$ &
\multicolumn{1}{c||}{$\lambda _{-}$} & Remarks \\ \hline\hline
$n<$ & $\mathbb{R}$ & $\mathbb{R}$ & $\mathbb{R}$ & $\mathbb{R}$ &
\multicolumn{1}{|c||}{$\mathbb{R}$} &  \\
$-1.30084$ & $>0$ & $<0$ & $>0$ & $>0$ & \multicolumn{1}{|c||}{$<0$} &
Unstable as $t \rightarrow 0$. \\ \hline
$-1.30084$ & $\mathbb{R}$ & $\mathbb{R}$ & $\mathbb{R}$ & $\mathbb{R}$ &
\multicolumn{1}{|c||}{$\mathbb{R}$} &  \\
$<n<$ & $>0$ & $<0$ & $>0$ & $<0$ & \multicolumn{1}{|c||}{$<0$} & Unstable
as $t \rightarrow 0$. \\
$-0.47942$ &  &  &  &  & \multicolumn{1}{|c||}{} &  \\ \hline
$-0.47942$ & $\mathbb{C}$ & $\mathbb{C}$ & $\mathbb{C}$ & $\mathbb{C}$ &
\multicolumn{1}{|c||}{$\mathbb{C}$} &  \\
$<n<$ & $\Re {(P_{+})}>0$ & $\Re {(\lambda _{1})}<0$ &  &  &
\multicolumn{1}{|c||}{} & Unstable as $t \rightarrow 0$. \\
$0.110873$ &  &  &  &  & \multicolumn{1}{|c||}{} &  \\ \hline
$0.110873$ & $\mathbb{R}$ & $\mathbb{R}$ & $\mathbb{R}$ & $\mathbb{C}$ &
\multicolumn{1}{|c||}{$\mathbb{C}$} &  \\
$<n<$ & $>0$ & $<0$ & $<0$ & $\Re {(\lambda _{+})}<0$ &
\multicolumn{1}{|c||}{$\Re {(\lambda _{-})}<0$} & Unstable as $t \rightarrow
0$. \\
$0.452692$ &  &  &  &  & \multicolumn{1}{|c||}{} &  \\ \hline
$0.452692$ & $\mathbb{R}$ & $\mathbb{R}$ & $\mathbb{R}$ & $\mathbb{R}$ &
\multicolumn{1}{|c||}{$\mathbb{R}$} &  \\
$<n<$ & $>0$ & $<0$ & $>0$ & $<0$ & \multicolumn{1}{|c||}{$<0$} & Unstable
as $t \rightarrow 0$. \\
$0.5$ &  &  &  &  & \multicolumn{1}{|c||}{} &  \\ \hline
$0.5$ & $\mathbb{R}$ & $\mathbb{R}$ & $\mathbb{R}$ & $\mathbb{R}$ &
\multicolumn{1}{|c||}{$\mathbb{R}$} &  \\
$<n<$ & $>0$ & $<0$ & $>0$ & $>0$ & \multicolumn{1}{|c||}{$<0$} & Unstable
as $t \rightarrow 0$. \\
$1$ &  &  &  &  & \multicolumn{1}{|c||}{} &  \\ \hline
& $\mathbb{R}$ & $\mathbb{R}$ & $\mathbb{R}$ & $\mathbb{R}$ &
\multicolumn{1}{|c||}{$\mathbb{R}$} &  \\
$n>1$ & $<0$ & $>0$ & $>0$ & $>0$ & \multicolumn{1}{|c||}{$<0$} &
Universe contracts. \\
&  &  &  &  & \multicolumn{1}{|c||}{} & $P_{+}<0$ \\ \hline
\end{tabular} \\

\bigskip

We note that for the solutions with $a(t)=t^{P_{+}/3}$, there is
always a negative eigenvalue, so are unstable for any $n$ as $t\rightarrow 0$%
. For $n>1$, $P_{+}<0$, so this corresponds to a contracting universe, in
which we are not interested here. However, we have $\Re {(P_{+})}>3$ for $1>n>%
\frac{1}{2}$ and $n<\frac{1-\sqrt{17}}{8}$, so we need to be careful that
the instability for these $n$ is not arising from the negative curvature
contribution characteristic of the Milne universe.  We expect the overall
assumption that the higher-order Ricci terms dominate the GR terms in the neighbourhood of the initial cosmological singularity to hold so long as $n>1/2$.

\subsection{Solutions with $P=P_{-}$}

Now consider the second case, with $P=P_{-}$, which turns out to be the most
physically relevant for consideration of the effects of higher-order ($n>1$)
corrections. We have

\begin{equation*}
P=\frac{3(1-3n+4n^{2})-\sqrt{3(-1+10n-5n^{2}-40n^{3}+48n^{4})}}{2(1-n)} ,
\end{equation*}%
with the stability decided by

\begin{eqnarray*}
\lambda _{1} &=&\frac{-20n^{2}+19n-5+\sqrt{3(-1+10n-5n^{2}-40n^{3}+48n^{4})}%
}{2(1-n)}, \\
\lambda _{2} &=&\frac{464n^{4}-392n^{3}+9n^{2}+64n-13-(36n^{2}-11n-3)\sqrt{%
3(-1+10n-5n^{2}-40n^{3}+48n^{4})}}{2(1-n)^{2}}.
\end{eqnarray*}

\begin{figure}[h]
\centering \includegraphics[scale=0.5]{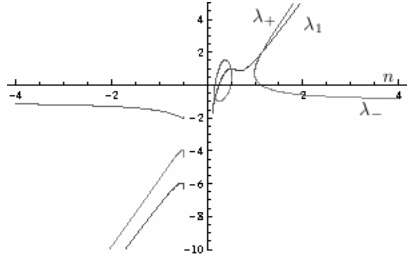}
\caption{Power-law exponents, $\lambda_{1}, \lambda_{\pm}$, versus $n$ for tensor perturbations with $%
P=P_{-}$.}
\label{fig:gminus}
\end{figure}

 \begin{tabular}{|c||c|c|c|c|c||r|}
\hline
\multicolumn{7}{|c|}{Values of $\lambda _{i}$ for different values of $n\neq
1$, taking $P=P_{-}$} \\
& $P_{-}$ & $\lambda _{1}$ & $\lambda _{2}$ & $\lambda _{+}$ &
\multicolumn{1}{c||}{$\lambda _{-}$} & Remarks \\ \hline\hline
$n<$ & $\mathbb{R}$ & $\mathbb{R}$ & $\mathbb{R}$ & $\mathbb{R}$ &
\multicolumn{1}{|c||}{$\mathbb{R}$} &  \\
$-0.47942$ & $>0$ & $<0$ & $>0$ & $<0$ & \multicolumn{1}{|c||}{$<0$} &
Unstable as $t \rightarrow 0$. \\ \hline
$-0.47942$ & $\mathbb{C}$ & $\mathbb{C}$ & $\mathbb{C}$ & $\mathbb{C}$ &
\multicolumn{1}{|c||}{$\mathbb{C}$} &  \\
$<n<$ & $\Re {(P_{-})}>0$ & $\Re {(\lambda _{1})}<0$ &  &  &
\multicolumn{1}{|c||}{} & Unstable as $t \rightarrow 0$. \\
$0.110873$ &  &  &  &  & \multicolumn{1}{|c||}{} &  \\ \hline
$0.110873$ & $\mathbb{R}$ & $\mathbb{R}$ & $\mathbb{R}$ & $\mathbb{C}$ &
\multicolumn{1}{|c||}{$\mathbb{C}$} &  \\
$<n<$ & $>0$ & $<0$ & $<0$ & $\Re {(\lambda _{+})}<0$ &
\multicolumn{1}{|c||}{$\Re {(\lambda _{-})}<0$} & Unstable as $t \rightarrow
0$. \\
$0.159452$ &  &  &  &  & \multicolumn{1}{|c||}{} &  \\ \hline
$0.159452$ & $\mathbb{R}$ & $\mathbb{R}$ & $\mathbb{R}$ & $\mathbb{R}$ &
\multicolumn{1}{|c||}{$\mathbb{R}$} &  \\
$<n<$ & $>0$ & $<0$ & $>0$ & $<0$ & \multicolumn{1}{|c||}{$<0$} & Unstable
as $t \rightarrow 0$. \\
$0.169938$ &  &  &  &  & \multicolumn{1}{|c||}{} &  \\ \hline
$0.169938$ & $\mathbb{R}$ & $\mathbb{R}$ & $\mathbb{R}$ & $\mathbb{R}$ &
\multicolumn{1}{|c||}{$\mathbb{R}$} &  \\
$<n<$ & $>0$ & $<0$ & $>0$ & $>0$ & \multicolumn{1}{|c||}{$<0$} & Unstable
as $t \rightarrow 0$. \\
$0.25$ &  &  &  &  & \multicolumn{1}{|c||}{} &  \\ \hline
$0.25$ & $\mathbb{R}$ & $\mathbb{R}$ & $\mathbb{R}$ & $\mathbb{R}$ &
\multicolumn{1}{|c||}{$\mathbb{R}$} & \\
$<n<$ & $<0$ & $>0$ & $>0$ & $>0$ & \multicolumn{1}{|c||}{$<0$} &
Universe contracts. \\
$0.5$ &  &  &  &  & \multicolumn{1}{|c||}{} & $P_{-} < 0$ \\ \hline
$0.5$ & $\mathbb{R}$ & $\mathbb{R}$ & $\mathbb{R}$ & $\mathbb{R}$ &
\multicolumn{1}{|c||}{$\mathbb{R}$} &  \\
$<n<$ & $>0$ & $>0$ & $>0$ & $>0$ & \multicolumn{1}{|c||}{$>0$} & Stable as $%
t \rightarrow 0$. \\
$0.520752$ &  &  &  &  & \multicolumn{1}{|c||}{} &  \\ \hline
$0.520752$ & $\mathbb{R}$ & $\mathbb{R}$ & $\mathbb{R}$ & $\mathbb{C}$ &
\multicolumn{1}{|c||}{$\mathbb{C}$} &  \\
$<n<$ & $>0$ & $>0$ & $<0$ & $\Re {(\lambda _{+})}>0$ &
\multicolumn{1}{|c||}{$\Re {(\lambda _{-})}>0$} & Stable as $t \rightarrow 0$%
. \\
$0.989666$ &  &  &  &  & \multicolumn{1}{|c||}{} &  \\ \hline
$0.989666$ & $\mathbb{R}$ & $\mathbb{R}$ & $\mathbb{R}$ & $\mathbb{R}$ &
\multicolumn{1}{|c||}{$\mathbb{R}$} &  \\
$<n<$ & $>0$ & $>0$ & $>0$ & $>0$ & \multicolumn{1}{|c||}{$>0$} & Stable as $%
t \rightarrow 0$. \\
$1.1309$ &  &  &  &  & \multicolumn{1}{|c||}{} &  \\ \hline
& $\mathbb{R}$ & $\mathbb{R}$ & $\mathbb{R}$ & $\mathbb{R}$ &
\multicolumn{1}{|c||}{$\mathbb{R}$} &  \\
$n>1.1309$ & $>0$ & $>0$ & $>0$ & $>0$ & \multicolumn{1}{|c||}{$<0$} &
Unstable as $t \rightarrow 0$. \\
&  &  &  &  & \multicolumn{1}{|c||}{} &  \\ \hline
\end{tabular}  \\

\bigskip

We saw that for $n>1$, $P_{+}<0$, so for an expanding universe, the only
relevant value for $P$ with $n>1$ is $P=P_{-}$. The first table shows that
this can only be stable as $t\rightarrow 0$ if
\begin{eqnarray}
\frac{1}{2}<n<\sqrt{2}\cos{\left[\frac{1}{3}\arccos{\left(\frac{-\sqrt{2}}{4}\right)}\right]}  \approx
1.1309.  \label{nstab}
\end{eqnarray}

In particular, for all integers $n>1$, the exact isotropic solution with $%
a(t)=t^{P_{-}/3}$ is \textit{not} a past attractor as $t\rightarrow 0$.
Thus, it appears that the quadratic ($n=1$) case studied earlier was
exceptional and the stability of the isotropic singularity found for that
case does not extend to higher-order corrections to general relativity with $%
n>1$.

\section{Vector (vortical) perturbations} \label{vector}
We have shown that for gravitational-wave perturbations there is a very
small range of values of $n$, given in eqn. (\ref{nstab}), for which the
perturbations are stable as $t\rightarrow 0$. We now consider the vortical
perturbations, which are of vector-type. The metric is:
\begin{equation*}
ds^{2}=-dt^{2}-2aB_{\alpha }dtdx^{\alpha }+a^{2}(\delta _{\alpha \beta
}+2C_{(\alpha ,\beta )})dx^{\alpha }dx^{\beta },
\end{equation*}%
where $B^{\alpha }\!_{,\alpha }\equiv 0\equiv C^{\alpha }\!_{,\alpha }$.

The energy-momentum tensor is decomposed as usual \cite{ellis},
\begin{equation}
T_{ab}=\mu u_{a}u_{b}+ph_{ab}+q_{a}u_{b}+q_{b}u_{a}+\pi _{ab},
\end{equation}%
where $h_{ab}\equiv g_{ab}+u_{a}u_{b},q_{a}u^{a}\equiv 0\equiv \pi
_{ab}u^{b} $ and $\pi _{a}^{a}\equiv 0$. The fluid four-velocity $u_{a}$ and
the energy flux $q_{a}$ are decomposed as (using $t$ as the time variable,
index `0' denotes $t$):
\begin{eqnarray}
u^{0}\equiv 1, &&u_{0}=-1,  \notag \\
u^{\alpha }\equiv a^{-1}V^{\alpha }, &&u_{\alpha }=a(V_{\alpha }-B_{\alpha }) ,
\notag \\
q_{0}=0, &&q_{\alpha }\equiv aQ_{\alpha } .
\end{eqnarray}

The energy-momentum tensor is then
\begin{eqnarray}
T^0_{\alpha}=(\mu + p)\left(V_{\alpha}+\frac{Q_{\alpha}}{\mu + p} -
B_{\alpha}\right), && \delta T^{\alpha}_{\beta}= \Pi^{\alpha}_{\beta} .
\label{em}
\end{eqnarray}

We can also decompose the perturbation variables as
\begin{eqnarray}
B_{\alpha }(\mathbf{x},t)\equiv b(t)Y_{\alpha }(\mathbf{x}), &C_{\alpha
}\equiv cY_{\alpha },&\Delta Y_{\alpha }\equiv -k^{2}Y_{\alpha } ,  \notag \\
V_{\alpha }\equiv vY_{\alpha }, &Q_{\alpha }\equiv qY_{\alpha },&\Pi _{\beta
}^{\alpha }\equiv p\pi _{T}Y_{\beta }^{\alpha }
\end{eqnarray}%
and introduce the gauge-invariant variables \cite{Noh:vort} :
\begin{eqnarray}
V_{\alpha }+\frac{Q_{\alpha }}{\mu +p}-B_{\alpha }=\left( v+\frac{q}{\mu +p}%
-b\right) Y_{\alpha } &\equiv &v_{\omega }Y_{\alpha }, \\
V_{\alpha }+\frac{Q_{\alpha }}{\mu +p}+C_{\alpha }^{\prime }=\left( v+\frac{q%
}{\mu +p}+c^{\prime }\right) Y_{\alpha } &\equiv &v_{\sigma }Y_{\alpha }, \\
B_{\alpha }+C_{\alpha }^{\prime }=(v_{\sigma }-v_{\omega })Y_{\alpha }
&\equiv &\Psi Y_{\alpha },
\end{eqnarray}%
where a prime denotes a derivative with respect to the conformal time
variable $\eta $; $v_{\omega }$ and $v_{\sigma }$ may be interpreted as the
velocity variables related to the vorticity and the shear respectively.

We will work in the ``C-gauge'', i.e. we set $C_\alpha = 0$, which
completely fixes the gauge condition. Then, using the quantities presented
in Appendix \ref{app:v}, we find that the perturbed parts of the tensor $P^a_b$ are:
\begin{eqnarray}
\delta P^0_0 &=&0 , \\
P^{0}_{\alpha} &=& nY^{n-1}\Biggl\{\frac{\Delta}{2a}\ddot{B}_\alpha+\frac{H}{%
2}\frac{\Delta}{a}\dot{B}_\alpha -(2\dot{H}+4H^2)\frac{\Delta}{a}B_\alpha -%
\frac{\Delta^2}{2a^3}B_\alpha \Biggr\}  \notag \\
&& +n(Y^{n-1})^{^{\textbf{.}}}\left\{\frac{\Delta}{2a}\dot{B}_\alpha+H\frac{\Delta}{a}%
B_\alpha \right\} , \\
\delta P^{\alpha}_{\beta} &=&-nY^{n-1}\frac{1}{a}\biggl\{\dddot{B}%
^{(\alpha}\,\!_{, \beta)}+3H\ddot{B}^{(\alpha}\,\!_{, \beta)}-\left(3\dot{H}%
+6H^2+\frac{\Delta}{a^2}\right)\dot{B}^{(\alpha}\,\!_{, \beta)}-4(\ddot{H}+6H\dot{H}+4H^3)B^{(\alpha}\,\!_{, \beta)} \biggr\}  \notag \\
&& -n(Y^{n-1})^{^{\textbf{.}}}\frac{1}{a}\left\{2\ddot{B}^{(\alpha}\,\!_{, \beta)}+5H%
\dot{B}^{(\alpha}\,\!_{, \beta)}-\left(2\dot{H}+4H^2+\frac{\Delta}{a^2}%
\right)B^{(\alpha}\,\!_{, \beta)}\right\}  \notag \\
&& -n(Y^{n-1})^{^{\textbf{..}}}\frac{1}{a}\left\{\dot{B}^{(\alpha}\,\!_{,
\beta)}+2HB^{(\alpha}\,\!_{, \beta)}\right\}  . \label{ab}
\end{eqnarray}

Combining the equations (\ref{em})-(\ref{ab}), we have
\begin{eqnarray}
T_{\alpha }^{0} &=&-\frac{\Delta B_{\alpha }}{2a}+A\Biggl[nY^{n-1}\Biggl\{%
\frac{\Delta }{2a}\ddot{B}_{\alpha }+\frac{1}{2}\left( (n-1)\frac{\dot{Y}}{Y}%
+H\right) \frac{\Delta }{a}\dot{B}_{\alpha }  \notag \\
&&+\left( (n-1)H\frac{\dot{Y}}{Y}-(2\dot{H}+4H^{2})\right) \frac{\Delta }{a}%
B_{\alpha }-\frac{\Delta ^{2}}{2a^{3}}B_{\alpha }\Biggr\}\Biggr]  \notag \\
&=&a(\mu +p)v_{\omega }Y_{\alpha }, \\
-\frac{1}{k}p\pi _{T} &=&\frac{2}{k^{2}a^{3}}\left[ a^{4}(\mu +p)v_{\omega }%
\right] ^{^{\textbf{.}}},
\end{eqnarray}%
and so for vanishing anisotropic pressure of the matter part, $p\pi _{T}=0$,
angular momentum is conserved exactly as in the quadratic case and $%
a^{3}T_{\alpha }^{0}\equiv \Omega Y_{\alpha }(\mathbf{x})$ is a constant in
time.

For $a=t^{P/3}$, $H=\frac{P}{3t}$, and $Y=\frac{4P^{2}}{9t^{4}}%
\left( 3-3P+P^{2}\right) \propto t^{-4}$, so we have therefore:

\begin{equation}
P^0_\alpha = nY^{n-1}\frac{\Delta}{2t^\frac{P}{3}}\Biggl\{\ddot{B}_\alpha+%
\frac{1}{t}\left(4+\frac{P}{3}-4n\right)\dot{B}_\alpha +\frac{1}{t^2}%
\left(4P-\frac{8P^2}{9}-\frac{8Pn}{3}\right)B_\alpha -\frac{\Delta}{t^\frac{%
2P}{3}}B_\alpha \Biggr\} .
\end{equation}

\subsection{The $A \rightarrow \infty$ limit}

\begin{figure}[h]
\centering \includegraphics[scale=0.5]{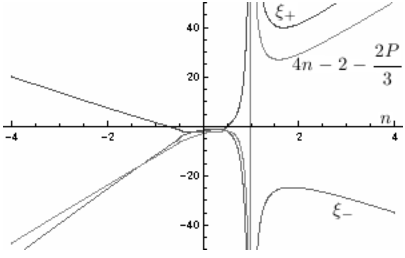}
\caption{Power-law exponents, $4n-2-\frac{2P}{3}, \xi_{\pm}$, versus $n$ for vector perturbations with $%
P=P_{+}$.}
\label{fig:vplus}
\end{figure}

\begin{figure}[h]
\centering \includegraphics[scale=0.5]{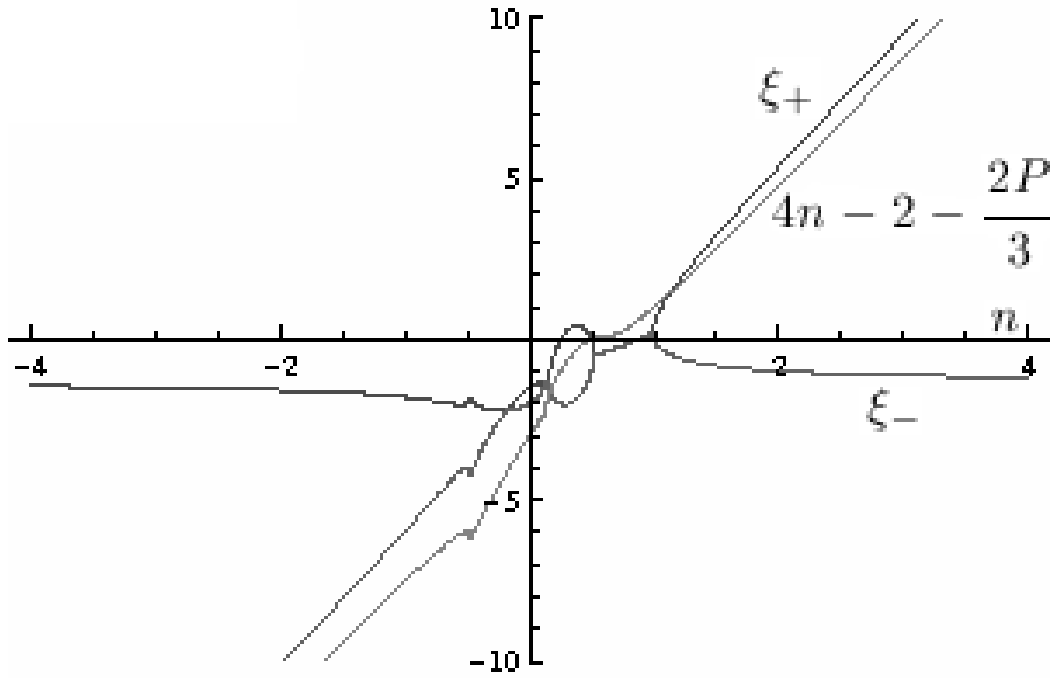}
\caption{Power-law exponents, $4n-2-\frac{2P}{3}, \xi_{\pm}$, versus $n$ for vector perturbations with $%
P=P_{-}$.}
\label{fig:vminus}
\end{figure}

In the limit $A\rightarrow \infty $, where the GR term can be neglected and
the higher-order Ricci terms dominate, we have
\begin{eqnarray}
\Omega \left( \frac{9}{4P^{2}(3-3P+P^{2})}\right) ^{n-1} &=&-nk^{2}t^{4-4n+%
\frac{2P}{3}}\Biggl\{\frac{\ddot{\Psi}}{2}+\frac{1}{2t}\left( -4(n-1)+\frac{P%
}{3}\right) \dot{\Psi}  \notag \\
&&+\frac{1}{t^{2}}\left( \frac{-4Pn}{3}+\frac{6P}{3}-\frac{4P^{2}}{9}\right)
\Psi +\frac{k^{2}}{2t^{\frac{2P}{3}}}\Psi \Biggr\} .
\end{eqnarray}
Defining the constant
\begin{equation*}
\tilde{\Omega}\equiv -\frac{2\Omega }{nk^{2}}\left( \frac{9}{%
4P^{2}(3-3P+P^{2})}\right) ^{n-1},
\end{equation*}
we have
\begin{equation*}
\tilde{\Omega}t^{4n-4-\frac{2P}{3}}=\ddot{\Psi}+\frac{1}{t}\left( -4(n-1)+%
\frac{P}{3}\right) \dot{\Psi}+\frac{2}{3t^{2}}\left( -4Pn+6P-\frac{4P^{2}}{3}%
\right) \Psi +\frac{k^{2}}{t^{\frac{2P}{3}}}\Psi .
\end{equation*}%
If we take the long-wavelength limit, i.e. we drop the last term on the
right-hand side, then we have to solve
\begin{equation*}
\tilde{\Omega}t^{4n-2-\frac{2P}{3}}=t^{2}\ddot{\Psi}+t\left( -4(n-1)+\frac{P%
}{3}\right) \dot{\Psi}+\frac{2}{3}\left( -4Pn+6P-\frac{4P^{2}}{3}\right) \Psi .
\end{equation*}%
The complementary function (LHS = 0) is solved by $\Psi =t^{\xi }$, where
\begin{eqnarray*}
0 &=&\xi ^{2}+\left( 3-4n+\frac{P}{3}\right) \xi +\frac{2}{3}\left( -4Pn+6P-%
\frac{4P^{2}}{3}\right)  \\
\Rightarrow \;\xi \;=\;\xi _{\pm }\!\!\! &\equiv &\frac{1}{6}\left( -9-P+12n\pm \sqrt{3(27-72n+48n^{2}-42P+24nP+11P^{2})}%
\right) , \label{xi} \\
&=&\frac{1}{6}\left( \lambda _{1}+8(n-1)\right) \pm \frac{1}{2}\sqrt{\lambda
_{2}},
\end{eqnarray*}%
where the $\lambda _{i}$ were defined in (\ref{lambda1}) and (\ref{lambda2}), and for stability as $%
t\rightarrow 0$, we need $\Re{(\xi _{\pm })}\geq 0$. The additional mode from the
particular solution has $\Psi \sim t^{4n-2-\frac{2P}{3}}=t^{\frac{2}{3}%
(\lambda _{1}+2(n-1))}$. The signs of these exponents for different values of $%
n$ are summarised in the tables which follow and their values are plotted in
figures \ref{fig:vplus} and \ref{fig:vminus} for $P_{+}$ and $P_{-}$
respectively.

\subsubsection{Solutions with $P=P_{+}$}

For $P=P_{+}$, we have $\Re{(\xi _{+})}\geq 0$ for $n\geq \frac{1}{2}$ and $n\leq -%
\frac{13}{18}$, whilst $\Re{(\xi _{-})}<0\;$for all $n$. Finally, from the
particular solution, the exponent $4n-2-\frac{2P}{3}$ is positive for $n>1$
and negative (and hence unstable as $t\rightarrow 0$) for $n<1$. Thus, for
the solution branch $P=P_{+}$, the vector modes are unstable as $t
\rightarrow 0$ for all values of $n$.

 \begin{tabular}{|c||c|c|c|c||r|}
\hline
\multicolumn{6}{|c|}{Values of $\xi_{i}$ for different values of $n\neq 1$,
taking $P=P_{+}$} \\
& $P_{+}$ & $\xi_{+}$ & $\xi_{-}$ & \multicolumn{1}{c||}{$4n-2-\frac{2P}{3}$}
& Remarks \\ \hline\hline
$n<-\frac{13}{18}$ & $\mathbb{R}$ & $\mathbb{R}$ & $\mathbb{R}$ &
\multicolumn{1}{|c||}{$\mathbb{R}$} &  \\
& $>0$ & $<0$ & $<0$ & \multicolumn{1}{|c||}{$<0$} & Unstable as $t
\rightarrow 0$. \\ \hline
$-\frac{13}{18}$ & $\mathbb{R}$ & $\mathbb{R}$ & $\mathbb{R}$ &
\multicolumn{1}{|c||}{$\mathbb{R}$} &  \\
$<n<$ & $>0$ & $>0$ & $<0$ & \multicolumn{1}{|c||}{$<0$} & Unstable as $t
\rightarrow 0$. \\
$-0.47942$ &  &  &  & \multicolumn{1}{|c||}{} &  \\ \hline
$-0.47942$ & $\mathbb{C}$ & $\mathbb{C}$ & $\mathbb{C}$ &
\multicolumn{1}{|c||}{$\mathbb{C}$} &  \\
$<n<$ & $\Re {(P_{+})}>0$ & $\Re {(\xi _{+})}<0$ & $\Re {(\xi _{-})}<0$ &
\multicolumn{1}{|c||}{$\Re {(4n-2-\frac{2P}{3})}<0$} & Unstable as $t
\rightarrow 0$. \\
$0.110873$ &  &  &  & \multicolumn{1}{|c||}{} &  \\ \hline
$0.110873$ & $\mathbb{R}$ & $\mathbb{R}$ & $\mathbb{R}$ &
\multicolumn{1}{|c||}{$\mathbb{R}$} &  \\
$<n<$ & $>0$ & $\Re {(\xi _{+})}<0$ & $<0$ & \multicolumn{1}{|c||}{$<0$} &
Unstable as $t \rightarrow 0$. \\
$0.5$ &  &  &  & \multicolumn{1}{|c||}{} &  \\ \hline
$0.5$ & $\mathbb{R}$ & $\mathbb{R}$ & $\mathbb{R}$ & \multicolumn{1}{|c||}{$%
\mathbb{R}$} &  \\
$<n<$ & $>0$ & $>0$ & $<0$ & \multicolumn{1}{|c||}{$<0$} & Unstable as $t
\rightarrow 0$. \\
$1$ &  &  &  & \multicolumn{1}{|c||}{} &  \\ \hline
& $\mathbb{R}$ & $\mathbb{R}$ & $\mathbb{R}$ & \multicolumn{1}{|c||}{$%
\mathbb{R}$} &  \\
$1<n$ & $<0$ & $>0$ & $<0$ & \multicolumn{1}{|c||}{$>0$} & Universe contracts. \\
&  &  &  & \multicolumn{1}{|c||}{} & $P_{+}< 0$ \\ \hline
\end{tabular}  \\ \bigskip

\subsubsection{Solutions with $P=P_{-}$}

For $P=P_{-}$, $\Re{(\xi_{+})}\geq 0$ for $\frac{1}{4}\leq n\leq \frac{1}{2}$ and $%
n\geq 0.861425$, $\Re{(\xi _{-})}\geq 0$ for $0.861425\leq n\leq 1$, whilst $4n-2-%
\frac{2P}{3}$ is positive for $n>0.5$ and negative for $n<0.5$. For this
branch, the vector perturbations are stable to linear order as $t
\rightarrow 0$ for \begin{equation}1 \geq n \geq \frac{1}{36}\left(25+2\sqrt{23}\sinh{\left[\frac{1}{3}\text{arcsinh}\left(\frac{316}{23\sqrt{23}}\right)\right]}\right) \approx 0.861425\end{equation} and are unstable for all other $n$.

 \begin{tabular}{|c||c|c|c|c||r|}
\hline
\multicolumn{6}{|c|}{Values of $\xi _{i}$ for different values of $n\neq 1$,
taking $P=P_{-}$} \\
& $P_{-}$ & $\xi_{+}$ & $\xi_{-}$ & \multicolumn{1}{c||}{$4n-2-\frac{2P}{3}$}
& Remarks \\ \hline\hline
$n<$ & $\mathbb{R}$ & $\mathbb{R}$ & $\mathbb{R}$ & \multicolumn{1}{|c||}{$%
\mathbb{R}$} &  \\
$-0.47942$ & $>0$ & $<0$ & $<0$ & \multicolumn{1}{|c||}{$<0$} & Unstable as $%
t \rightarrow 0$. \\ \hline
$-0.47942$ & $\mathbb{C}$ & $\mathbb{C}$ & $\mathbb{C}$ &
\multicolumn{1}{|c||}{$\mathbb{C}$} &  \\
$<n<$ & $\Re {(P_{-})}>0$ & $\Re {(\xi _{+})}<0$ & $\Re {(\xi _{-})}<0$ &
\multicolumn{1}{|c||}{$\Re {(4n-2-\frac{2P}{3})}<0$} & Unstable as $t
\rightarrow 0$. \\
$0.110873$ &  &  &  & \multicolumn{1}{|c||}{} &  \\ \hline
$0.110873$ & $\mathbb{R}$ & $\mathbb{C}$ & $\mathbb{C}$ &
\multicolumn{1}{|c||}{$\mathbb{R}$} &  \\
$<n<$ & $>0$ & $\Re {(\xi _{+})}<0$ & $\Re {(\xi _{-})}<0$ &
\multicolumn{1}{|c||}{$<0$} & Unstable as $t \rightarrow 0$. \\
$0.159452$ &  &  &  & \multicolumn{1}{|c||}{} &  \\ \hline
$0.159452$ & $\mathbb{R}$ & $\mathbb{R}$ & $\mathbb{R}$ &
\multicolumn{1}{|c||}{$\mathbb{R}$} &  \\
$<n<$ & $>0$ & $<0$ & $<0$ & \multicolumn{1}{|c||}{$<0$} & Unstable as $t
\rightarrow 0$. \\
$0.25$ &  &  &  & \multicolumn{1}{|c||}{} &  \\ \hline
$0.25$ & $\mathbb{R}$ & $\mathbb{R}$ & $\mathbb{R}$ & \multicolumn{1}{|c||}{$%
\mathbb{R}$} &  \\
$<n<$ & $<0$ & $>0$ & $<0$ & \multicolumn{1}{|c||}{$<0$} & Universe contracts. \\
$0.5$ &  &  &  & \multicolumn{1}{|c||}{} & $P_{-}<0$ \\ \hline
$0.5$ & $\mathbb{R}$ & $\mathbb{R}$ & $\mathbb{R}$ & \multicolumn{1}{|c||}{$%
\mathbb{R}$} &  \\
$<n<$ & $>0$ & $<0$ & $<0$ & \multicolumn{1}{|c||}{$>0$} & Unstable as $t
\rightarrow 0$. \\
$0.520752$ &  &  &  & \multicolumn{1}{|c||}{} &  \\ \hline
$0.520752$ & $\mathbb{R}$ & $\mathbb{C}$ & $\mathbb{C}$ &
\multicolumn{1}{|c||}{$\mathbb{R}$} &  \\
$<n<$ & $>0$ & $\Re {(\xi _{+})}<0$ & $\Re {(\xi _{-})}<0$ &
\multicolumn{1}{|c||}{$>0$} & Unstable as $t \rightarrow 0$. \\
$0.861425$ &  &  &  & \multicolumn{1}{|c||}{} &  \\ \hline
$0.861425$ & $\mathbb{R}$ & $\mathbb{C}$ & $\mathbb{C}$ &
\multicolumn{1}{|c||}{$\mathbb{R}$} &  \\
$<n<$ & $>0$ & $\Re {(\xi _{+})}>0$ & $\Re {(\xi _{-})}>0$ &
\multicolumn{1}{|c||}{$>0$} & Stable as $t \rightarrow 0$. \\
$0.989666$ &  &  &  & \multicolumn{1}{|c||}{} &  \\ \hline
$0.989666$ & $\mathbb{R}$ & $\mathbb{R}$ & $\mathbb{R}$ &
\multicolumn{1}{|c||}{$\mathbb{R}$} &  \\
$<n<$ & $>0$ & $>0$ & $>0$ & \multicolumn{1}{|c||}{$>0$} & Stable as $t
\rightarrow 0$. \\
$1$ &  &  &  & \multicolumn{1}{|c||}{} &  \\ \hline
& $\mathbb{R}$ & $\mathbb{R}$ & $\mathbb{R}$ & \multicolumn{1}{|c||}{$%
\mathbb{R}$} &  \\
$1<n$ & $>0$ & $>0$ & $<0$ & \multicolumn{1}{|c||}{$>0$} & Unstable as $t
\rightarrow 0$. \\
&  &  &  & \multicolumn{1}{|c||}{} &  \\ \hline
\end{tabular}  \\ \bigskip

\newpage \section{Scalar perturbations} \label{scalar}

We will now consider scalar perturbations. The metric for the general
scalar-type perturbation takes the form
\begin{eqnarray}
ds^{2}&=&-(1+2\alpha )dt^{2}-2a\beta _{,\alpha }dtdx^{\alpha } +a^{2}\left(
\delta _{\alpha \beta }(1+2\phi )+2\gamma _{,\alpha \beta }\right)
dx^{\alpha }dx^{\beta }.
\end{eqnarray}%
We use the proper time, $t$, as the time variable and also define the quantities, $\chi \equiv
a(\beta +a\dot{\gamma})$, $f\equiv Y^{n-1}\equiv (R_{ab}R^{ab})^{n-1}$. We
use overbars and deltas to denote background and perturbed quantities, so
that in general, $A=\overline{A}+\delta A$, and in particular $\overline{f}=%
\overline{Y}^{n-1},\delta f=(n-1)\overline{Y}^{n-2}\delta Y$. The important
quantities to linear order in the perturbation are given in Appendix \ref{app:s}.

Using the gravitational field equations, we obtain the complete set of
gauge-ready equations for the perturbed variables: \\

Energy:
\begin{eqnarray*}
\delta P^0_0 &=& n\overline{f}\biggl[-\frac{1}{2}\delta Y +12H^2\ddot{\alpha%
}+(12H\dot{H}+18H^3)\dot{\alpha}  +12(4H\ddot{H}-4\dot{H}^2 +6H^2\dot{H}-6H^4)\alpha \notag \\
&& -(16\dot{H}+34H^2)%
\frac{\Delta}{a^2}\alpha -2\frac{\Delta^2}{a^4}\alpha-12H\dddot{\phi}+(24\dot{H}-18H^2)\ddot{\phi}-(12\ddot{H}+36H\dot{H}-72H^3)%
\dot{\phi} \notag \\
&& +4\frac{\Delta}{a^2}\ddot{\phi}+24H\frac{\Delta}{a^2}\dot{\phi} -(4%
\dot{H}+20H^2)\frac{\Delta}{a^2}\phi  -2\frac{\Delta^2}{a^4}\phi  -4H\frac{\Delta}{a^2}\ddot{\chi}+(8\dot{H}%
+10H^2)\frac{\Delta}{a^2}\dot{\chi}  \notag \\
&&-(4\ddot{H}+20H\dot{H}-20H^3)\frac{\Delta%
}{a^2}\chi  +2\frac{\Delta^2}{a^4}\dot{\chi} +2H\frac{\Delta^2}{a^4}\chi
\biggr]  \notag \\
&& +n\dot{\overline{f}} \biggl[12H^2\dot{\alpha}+24H(2\dot{H}+3H^2)\alpha+4H%
\frac{\Delta}{a^2}\alpha-12H\ddot{\phi}-(12\dot{H}+54H^2)\dot{\phi}+4H\frac{%
\Delta}{a^2}\phi \\
&& -4H\frac{\Delta}{a^2}\dot{\chi} -(4\dot{H}+10H^2)\frac{\Delta}{a^2}\chi%
\biggr]  \notag \\
&& +n\biggl[(-12H\ddot{H} +12\dot{H}^2 -18H^2\dot{H}+18H^4)\delta f -3H(4%
\dot{H}+6H^2)\dot{\delta f} +(4\dot{H}+6H^2)\frac{\Delta}{a^2} \delta f %
\biggr]
\end{eqnarray*}

Momentum:
\begin{eqnarray*}
P^0_\alpha &=& n\overline{f}\Biggl[-4H\ddot{\alpha}-(12\dot{H}+12H^2)\dot{%
\alpha}+(-12\ddot{H}-48H\dot{H})\alpha -2 \frac{\Delta}{a^2}\dot{\alpha}+4H%
\frac{\Delta}{a^2}\alpha \\
&& +4\dddot{\phi}+12H\ddot{\phi}+(24\dot{H})\dot{\phi}-2\frac{\Delta}{a^2}%
\dot{\phi}+8H\frac{\Delta}{a^2}\phi+2\frac{\Delta}{a^2}\ddot{\chi}-2H\frac{%
\Delta}{a^2}\dot{\chi}+(2\dot{H}-8H^2)\frac{\Delta}{a^2}\chi \Biggr]%
_{,\alpha} \\
&& +n\dot{\overline{f}} \left[-4H\dot{\alpha}-(12\dot{H}+18H^2)\alpha-2
\frac{\Delta}{a^2}\alpha +4\ddot{\phi}+12H\dot{\phi} -2\frac{\Delta}{a^2}%
\phi+2 \frac{\Delta}{a^2}\dot{\chi} +2H\frac{\Delta}{a^2}\chi \right]%
_{,\alpha} \\
&& +n(4\dot{H}+6H^2)\dot{\delta f}_{,\alpha} +n(4\ddot{H}+12H\dot{H}-6H^3)
\delta f_{,\alpha}
\end{eqnarray*}

Trace:
\begin{eqnarray*}
\delta P^\alpha_\alpha &=& n\ddot{\overline{f}} \Biggl[12H\dot{\alpha}+3(16%
\dot{H}+24H^2)\alpha + 4\frac{\Delta}{a^2}\alpha -12\ddot{\phi}-36H\dot{\phi}%
+4\frac{\Delta}{a^2}\phi -4\frac{\Delta}{a^2}\dot{\chi}-4H\frac{\Delta}{a^2}%
\chi \Biggr]  \notag \\
&& +n\dot{\overline{f}} \Biggl[ 24H\ddot{\alpha}+3(28\dot{H}+42H^2)\dot{%
\alpha}+3(32\ddot{H}+144H\dot{H}+48H^3)\alpha+8 \frac{\Delta}{a^2}\dot{\alpha%
}-12H \frac{\Delta}{a^2}\alpha  \notag \\
&& -24\dddot{\phi}-108H\ddot{\phi}-108(\dot{H}+H^2)\dot{\phi}+16 \frac{\Delta%
}{a^2}\dot{\phi}-16H \frac{\Delta}{a^2}\phi  \notag \\
&& -8\frac{\Delta}{a^2}\ddot{\chi}-4H \frac{\Delta}{a^2}\dot{\chi}-(20\dot{H}%
-4H^2) \frac{\Delta}{a^2}\chi \Biggr] \\
&& +n\overline{f}\Biggl[12H\dddot{\alpha}+3(16\dot{H}+24H^2)\ddot{\alpha}%
+3(24\ddot{H}+108H\dot{H}+36H^3)\dot{\alpha}  \notag \\
&& +3(16\dddot{H}+96H\ddot{H}+72\dot{H}^2+144H^2\dot{H})\alpha+4\frac{\Delta%
}{a^2}\ddot{\alpha} -16H\frac{\Delta}{a^2}\dot{\alpha}-(20\dot{H}+34H^2)%
\frac{\Delta}{a^2}\alpha -2\frac{\Delta^2}{a^4}\alpha \\
&& -12\ddddot{\phi}-72H\dddot{\phi} -3(36\dot{H}+36H^2)\ddot{\phi}-3(24\ddot{%
H}+72H\dot{H})\dot{\phi}  \notag \\
&& +16\frac{\Delta}{a^2}\ddot{\phi}+16H\frac{\Delta}{a^2}\dot{\phi}-(16\dot{H%
}+8H^2)\frac{\Delta}{a^2}\phi-6\frac{\Delta^2}{a^4}\phi  \notag \\
&& -4\frac{\Delta}{a^2}\dddot{\chi}-(12\dot{H}-12H^2)\frac{\Delta}{a^2}\dot{%
\chi}-(16\ddot{H}-8H^3)\frac{\Delta}{a^2}\chi+2\frac{\Delta^2}{a^4}\dot{\chi}%
+6H\frac{\Delta^2}{a^4}\chi \Biggr] \\
&& -3n\Biggl[\ddot{\delta f}(4\dot{H}+6H^2) + \dot{\delta f}(8\ddot{H}+36H%
\dot{H} +12H^3) + \delta f (4\dddot{H}+24H\ddot{H}+12\dot{H}^2+18H^2\dot{H}%
-18H^4) \Biggr] \\
&& +4n(\dot{H}+3H^2) \frac{\Delta}{a^2} \delta f
\end{eqnarray*}

Trace-free:
\begin{eqnarray*}
P_{\beta }^{\alpha }-\frac{1}{3}\delta _{\beta }^{\alpha }P_{\alpha
}^{\alpha } &=&\frac{n}{a^{2}}\left( \nabla ^{\alpha }\nabla _{\beta }-\frac{%
1}{3}\delta _{\beta }^{\alpha }\Delta \right) \Biggl[\ddot{\overline{f}}%
\left[ \alpha +\phi -\dot{\chi}-H\chi \right] \\
&&+\dot{\overline{f}}\left[ 2\dot{\alpha}+3H\alpha -2\dot{\phi}-H\phi -2%
\ddot{\chi}-H\dot{\chi}+(4\dot{H}+7H^{2})\chi \right] \\
&&+\overline{f}\biggl[\ddot{\alpha}+5H\dot{\alpha}+(4\dot{H}+8H^{2})\alpha +%
\frac{\Delta }{a^{2}}\alpha -5\ddot{\phi}-17H\dot{\phi}-(4\dot{H}%
+8H^{2})\phi +3\frac{\Delta }{a^{2}}\phi \\
&&-\dddot{\chi}+(6\dot{H}+9H^{2})\dot{\chi}+(5\ddot{H}+21H\dot{H}%
+8H^{3})\chi -\frac{\Delta }{a^{2}}\dot{\chi}-3H\frac{\Delta }{a^{2}}\chi
\biggr] \\
&&-2(\dot{H}+3H^{2})\delta f\Biggr],
\end{eqnarray*}

with
\begin{eqnarray}
\delta Y &=& -12H(2\dot{H}+3H^2)\dot{\alpha}-48(\dot{H}^2+3H^2\dot{H}+3H^4)
\alpha -4(2\dot{H}+3H^2)\frac{\Delta}{a^2}\alpha +12(2\dot{H}+3H^2)\ddot{\phi%
}  \notag \\
&&+72H(\dot{H}+2H^2)\dot{\phi} -8(\dot{H}+3H^2)\frac{\Delta}{a^2}\phi +4(2%
\dot{H}+3H^2)\frac{\Delta}{a^2}\dot{\chi}+8H(\dot{H}+3H^2)\frac{\Delta}{a^2}%
\chi
\end{eqnarray}

The special unperturbed solution has $a=t^{P/3}, H=\frac{P}{3t}$ and  $\overline{Y}=\frac{4P^{2}}{9t^{4}}\left( 3-3P+P^{2}\right) \propto t^{-4}$.

\subsection{Linearising about the special solution in the zero-shear gauge}

We now take the large-scale limit and choose the zero-shear gauge ($\chi
\equiv 0$) and linearise about the special flat FRW solution with $a=t^{P/3}$. For $P=P_{\pm }$, the equations simplify to:

Energy:
\begin{eqnarray*}
\delta P^0_0 &=& n\left(\frac{4P^2}{9t^4}(3-3P+P^2)\right)^{n-2} \Biggl\{%
\frac{8P^4}{27t^7}\left(-6+6P-P^2+n(12-12P+3P^2)\right)\Biggr\} \times \\
&& \Biggl\{t\ddot{\alpha}+(2-4n+P)\dot{\alpha} - \frac{3}{P}\left(t^2\dddot{%
\phi} + (5-4n+P)t\ddot{\phi} + (4-8n+2P)\dot{\phi}\right) \Biggr\}
\end{eqnarray*}

Momentum:
\begin{eqnarray*}
\delta P^0_{\alpha} &=& n\left(\frac{4P^2}{9t^4}(3-3P+P^2)\right)^{n-2} %
\Biggl\{ \frac{8P^3}{27t^6}\left(6-6P+P^2-n(12-12P+3P^2)\right)\Biggr\} %
\times \\
&& \nabla_{\alpha} \Biggl\{t\ddot{\alpha} + (1-4n+P)\dot{\alpha} -\frac{3}{P}
\left(t^2 \dddot{\phi} + (4-4n+P)t\ddot{\phi} + (2-8n+2P)\dot{\phi} \right) %
\Biggr\}
\end{eqnarray*}

Trace:
\begin{eqnarray*}
\delta T^{\alpha}_{\alpha} &=& n\left(\frac{4P^2}{9t^4}(3-3P+P^2)%
\right)^{n-2}\Biggl[ \frac{8P^3}{9t^7}(-6+6P-P^2+3n(P-2)^2)\Biggr] \times \\
&& \Biggl[t^2\dddot{\alpha} +(4-8n+2P)t\ddot{\alpha}+(2-4n+P)(1-4n+P)\dot{%
\alpha} \\
&& -\frac{3}{P}\left(t^3\ddddot{\phi} + (8-8n+2P)t^2\dddot{\phi} +
(7-4n+P)(2-4n+P)t\ddot{\phi} + 2(2-4n+P)(1-4n+P)\dot{\phi}\right) \Biggl]
\end{eqnarray*}

Trace-free propagation:
\begin{eqnarray*}
\delta T_{\beta }^{\alpha }-\frac{1}{3}\delta _{\beta }^{\alpha }\delta
T_{\alpha }^{\alpha } &=&n\frac{1}{a^{2}}\left( \frac{4P^{2}}{9t^{4}}%
(3-3P+P^{2})\right) ^{n-2}\left( \nabla ^{\alpha }\nabla _{\beta }-\frac{1}{3%
}\delta _{\beta }^{\alpha }\Delta \right) \Biggl[\frac{4P^{2}}{9t^{4}}%
(3-3P+P^{2})\ddot{\alpha} \\
&&+\frac{4P^{2}}{27t^{5}}(72-69P+27P^{2}-P^{3}+6n(-12+14P-7P^{2}+P^{3}))\dot{%
\alpha} \\
&&+\frac{16P^{2}}{81t^{6}}%
(3-3P+P^{2})(27+12P-4P^{2}+3n(-21-5P+2P^{2})+36n^{2})\alpha \\
&&+\frac{4P^{2}}{9t^{4}}(-3-3P+P^{2}+6n(-2+3P-P^{2}))\ddot{\phi} \\
&&+\frac{4P^{2}}{27t^{5}}(-72+57P-33P^{2}+7P^{3}+12n(2-P)(3-3P+2P^{2}))\dot{%
\phi} \\
&&+\frac{16P^{2}}{81t^{6}}(3-3P+P^{2})(27-2P^{2}+n(-63+3P)+36n^{2})\phi %
\Biggr] .
\end{eqnarray*}

The energy and momentum equations together imply that
\begin{equation}
\alpha = \frac{3}{P}(t \dot{\phi}+\phi+\alpha_0) ,
\end{equation}
where $\alpha_0$ is a free constant. In addition, this also satisfies the
trace equation.

\begin{figure}[h]
\centering \includegraphics[scale=0.5]{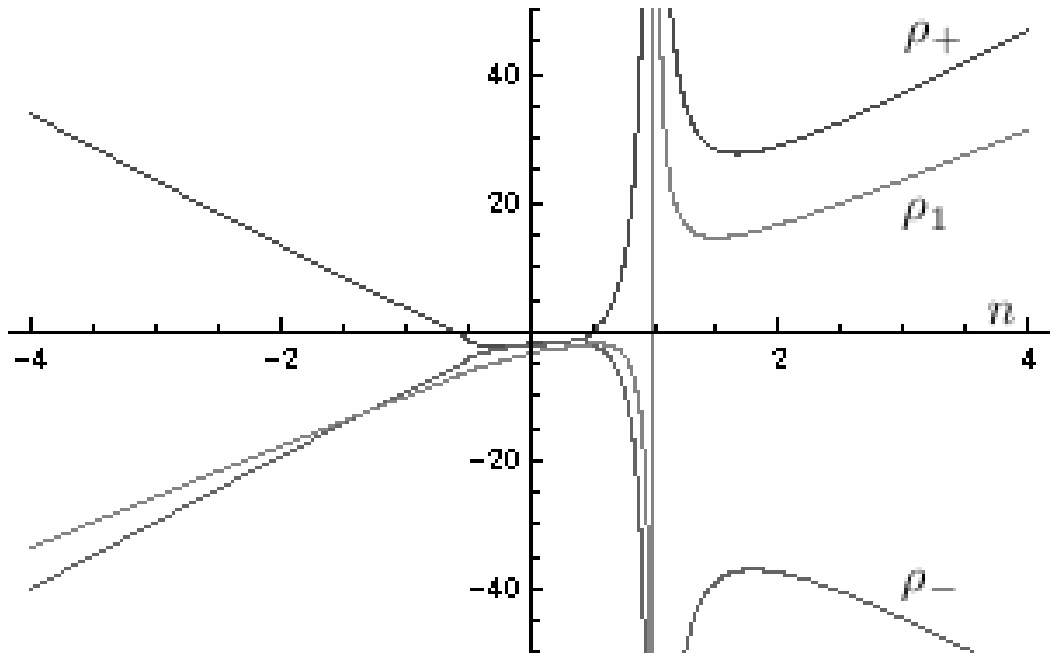}
\caption{Power-law exponents, $\rho_{1}, \rho_{\pm}$, versus $n$ for scalar perturbations with $%
P=P_{+}$.}
\label{fig:splus}
\end{figure}

\begin{figure}[h]
\centering \includegraphics[scale=0.5]{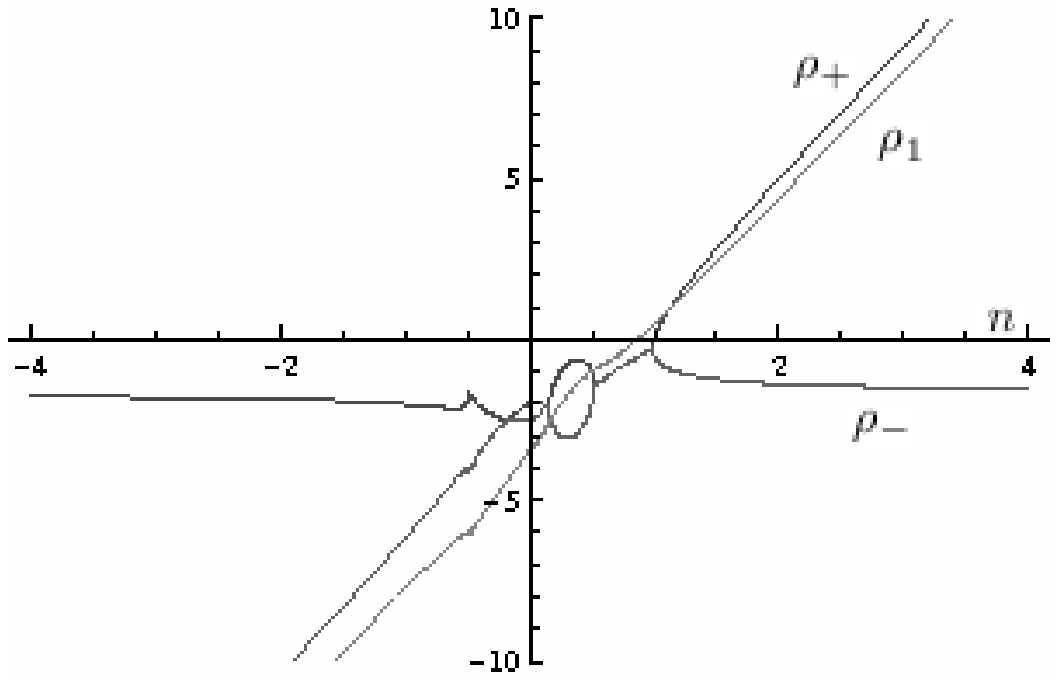}
\caption{Power-law exponents, $\rho_{1}, \rho_{\pm}$, versus $n$ for scalar perturbations with $%
P=P_{-}$.}
\label{fig:sminus}
\end{figure}

Finally, we use the trace-free equation to calculate that
\begin{eqnarray}
\phi &=&\phi _{0}+\phi _{1}t^{\rho _{1}}+\phi _{+}t^{\rho _{+}}+\phi
_{-}t^{\rho _{-}} , \\
\alpha &=&\frac{3}{P}\left[ (1+\rho _{1})\phi _{1}t^{\rho _{1}}+(1+\rho
_{+})\phi _{+}t^{\rho _{+}}+(1+\rho _{-})\phi _{-}t^{\rho _{-}}\right]
\notag \\
&&-\frac{27-2P^{2}+n(-63+3P)+36n^{2}}{27+12P-4P^{2}+3n(-21-5P+2P^{2})+36n^{2}%
}\phi _{0} , \\
\rho _{1} &\equiv &-3+4n-\frac{P}{3} , \\
\rho _{\pm } &\equiv &\frac{1}{6}\left( -15+12n+P\pm \sqrt{%
3(27-42P+11P^{2}-72n+24nP+48n^{2})}\right)  \notag \\
&=&\xi _{\pm }+\frac{P}{3}-1,
\end{eqnarray}%
where $\phi _{0},\phi _{1}$ and $\phi _{\pm }$ are free constants, and $\xi
_{\pm }$ were defined in (\ref{xi}). We note that these are the same
power-law exponents as for the vorticity perturbations plus $\frac{P}{3}-1$. These
exponents are shown in figures \ref{fig:splus} and \ref{fig:sminus}. For all
$n$, and either choice of $P=$ $P_{\pm }$, at least one of these exponents
has negative real part, and hence the isotropic vacuum solution is unstable as $%
t\rightarrow 0$.

\bigskip

\section{Summary} \label{summary}

We can now summarise the results for the linearised tensor, vector and
scalar perturbations about the spatially flat vacuum FRW solution (\ref{nsoln}) of the theory with Lagrangian $(R_{ab}R^{ab})^{n}$. The general
perturbed metric in the neighbourhood of the isotropic vacuum solution $%
a=t^{P/3},$ given by equation (\ref{nsoln}), has the form
\begin{eqnarray}
ds^{2}&=&-(1+2\alpha )dt^{2}-a\tilde{B}_{\alpha }dtdx^{\alpha } +a^{2}(\delta
_{\alpha \beta }+\tilde{C}_{\alpha \beta })dx^{\alpha }dx^{\beta },
\end{eqnarray}%
and the perturbation variables may be decomposed into their scalar, vector
and tensor parts by writing
\begin{eqnarray*}
\tilde{B}_{\alpha } &=&2\beta _{,\alpha }+2B_{\alpha }, \\
\tilde{C}_{\alpha \beta } &=&2\phi \delta _{\alpha \beta }+2\gamma _{,\alpha
\beta }+2C_{(\alpha ,\beta )}+2C_{\alpha \beta }.
\end{eqnarray*}
In the gauge defined by $\beta \equiv 0 \equiv \gamma$ and $C_{\alpha} \equiv 0$, the
general solution of the linearised equations is given by
\begin{eqnarray}
C^{\alpha}_{\beta} &=& a^{\alpha}_{\beta}(\mathbf{x})+t^{\lambda_{1}}b^{%
\alpha}_{\beta}(\mathbf{x})+t^{\lambda_{+}}c^{\alpha}_{\beta}(\mathbf{x}%
)+t^{\lambda_{-}}d^{\alpha}_{\beta}(\mathbf{x}) \\
B_{\alpha} &=& t^{\xi_{+}}Y^{(1)}_{\alpha}(\mathbf{x})+t^{\xi_{-}}Y^{(2)}_{\alpha}(\mathbf{x})  -\frac{%
3\tilde{\Omega} t^{4n-2-\frac{2P}{3}}}{2(3-6P+P^2-6n+6nP)} \\
\phi &=& \phi_{0}(\mathbf{x}) + \phi_{1}(\mathbf{x})t^{-3+4n-\frac{P}{3}} + \phi_{+}(\mathbf{x})t^{\xi_{+}-1+%
\frac{P}{3}} + \phi_{-}(\mathbf{x})t^{\xi_{-}-1+\frac{P}{3}} \\
\alpha & = & -\alpha_0 \phi_{0} -\frac{6-12n+P}{P} \phi_{1}t^{-3+4n-\frac{P}{%
3}}  + \left(\frac{3\xi_{+}}{P}+1 \right)\phi_{+}t^{\xi_{+}-1+\frac{P}{3}} +
\left(\frac{3\xi_{-}}{P}+1 \right)\phi_{-}t^{\xi_{-}-1+\frac{P}{3}},
\end{eqnarray}
where
\begin{eqnarray*}
\lambda_{1} & = & -1-P+4n \\
\lambda_{2} & = & 9-14P+\frac{11}{3}P^2-24n+8Pn+16n^2 \\
\lambda_{\pm} &=& \frac{1}{2}\left(\lambda_{1} \pm \sqrt{\lambda_{2}}\right)
\\
\xi_{\pm} &=& \frac{1}{6}(-9-P-12n)\pm \frac{1}{2} \sqrt{\lambda_{2}} \\
&=& \frac{1}{6}(\lambda_{1}+8(n-1))\pm \frac{1}{2} \sqrt{\lambda_{2}} \\
\alpha_0 &=& -\frac{27-2P^2+n(-63+3P)+36n^2}{%
27+12P-4P^2+3n(-21-5P+2P^2)+36n^2}
\end{eqnarray*}

For the solution
branch defined by $P=P_{+}$, each type of perturbation (tensor, vector and scalar) is unstable as $t \rightarrow 0$ for all values of $n$. For $P=P_{-}$, which is the only physically relevant value of $P$ for $n > 1$, the tensor perturbations are stable to linear order as $t \rightarrow 0$ for \begin{equation*}
\frac{1}{2}<n<\sqrt{2}\cos{\left[\frac{1}{3}\arccos{\left(\frac{-\sqrt{2}}{4}\right)}\right]}  \approx
1.1309
\end{equation*} and the vector perturbations are stable to linear order as $t
\rightarrow 0$ for \begin{equation*}1 \geq n \geq \frac{1}{36}\left(25+2\sqrt{23}\sinh{\left[\frac{1}{3}\text{arcsinh}\left(\frac{316}{23\sqrt{23}}\right)\right]}\right) \approx 0.861425 .\end{equation*} For all other $n$, these perturbations are unstable as $t \rightarrow 0$. The scalar perturbations are unstable as $t \rightarrow 0$ for all $n$.

In conclusion, in our earlier work \cite{mid} we discovered that isotropic
cosmological models in theories of gravity formed with a quadratic Ricci
term added to the Einstein-Hilbert action are stable on approach to an
initial `Big Bang' singularity. This is quite different to the behaviour of
general relativistic cosmological models, where isotropy is strongly
unstable in this limit in vacuum \cite{misner, bkl}. In this paper, we have analysed the more
complicated problem of cosmological evolution in the presence of arbitrary
powers of the Ricci term in the Lagrangian. We have found that the
behaviour displayed in the quadratic case was special. Isotropic power-law
solutions of the sort found by Barrow and Clifton \cite{Clifton2, Clifton3} still
exist in vacuum and with a perfect fluid when a term proportional to $%
(R_{ab}R^{ab})^{n}$ is added to the Einstein-Hilbert action for general $%
n\neq 1$. However, both solution branches of these special isotropic
solutions are unstable to the growth of small metric perturbations as $%
t\rightarrow 0$, and so the quadratic case with $n=1$, in which these
perturbations are bounded in this limit, is special.

\textbf{Acknowledgements}:
Jonathan Middleton acknowledges a PPARC/STFC studentship. We would like to
thank Timothy Clifton, Sigbj\o rn Hervik and Kei-ichi Maeda for helpful
discussions. \ \ \ \ \ \ \ \ \ \ \ \ \ \ \ \ \ \ \ \ \ \ \ \ \ \ \ \ \

\appendix

\section{Relevant tensor quantities using t as the time variable} \label{app:t}

Using $t$ as the time variable and $H=\frac{\dot{a}}{a}$, the important quantities to linear order in the gravitational wave-type
perturbation are:
\begin{eqnarray*}
\Gamma^{0}\!_{\alpha \beta}&=& a^2[H(\delta_{\alpha \beta} +2C_{\alpha
\beta}) +\dot{C}_{\alpha \beta}] \\
\Gamma^{\alpha}\!_{0 \beta}&=& H \delta^{\alpha}_{\beta} +\dot{C}%
^{\alpha}_{\beta} \\
\Gamma^{\alpha}\!_{\beta \gamma}&=& C^{\alpha}\!_{\beta , \gamma} +
C^{\alpha}\!_{\gamma , \beta} - C_{\gamma \beta ,}\! ^{\alpha} \\
\Gamma &=& 0 \; \text{otherwise} \\
g^{cd}\Gamma^{0}\!_{cd}&=& 3H \\
R_0^0 &=& 3(\dot{H} +H^2) \\
R_0^{\alpha} &=& R^0_{\alpha} = 0 \\
R^{\alpha}_{\beta} &=& (\dot{H} +3H^2)\delta^{\alpha}_{\beta} +\ddot{C}%
^{\alpha}_{\beta}+3H\dot{C}^{\alpha}_{\beta}-\frac{\Delta}{a^2}%
C^{\alpha}_{\beta} \\
R &=& 6(\dot{H}+2H^2) \\
Y\equiv R^a_bR^b_a &=& 12(\dot{H}^2 +3\dot{H}H^2 +3H^4) \\
R^{cd}R^{0}\!_{cd0} &=& -3(\dot{H}+H^2)(\dot{H}+3H^2) \\
R^{cd}R^{0}\!_{cd\alpha} & =& 0 \, = \, R^{cd}R^{\alpha}\!_{cd0} \\
R^{cd}R^{\alpha}\!_{cd\beta} &=& -(3\dot{H}^2+8\dot{H}H^2+9H^4)\delta^{%
\alpha}_{\beta}-(3\dot{H}+2H^2)\ddot{C}^{\alpha}_{\beta}  -(7\dot{H}+6H^2)H\dot{C}^{\alpha}_{\beta}\notag \\
&& +(\dot{H}+2H^2)\frac{\Delta}{a^2}%
C^{\alpha}_{\beta} \\
\nabla _{0}R^{0}_{0} &=& 3(\ddot{H} +2H\dot{H}) \\
\nabla _{0}R_{0}^{\alpha} &=&0 \, = \, \nabla _{0}R_{\alpha}^{0} \, = \,
\nabla _{\alpha}R_{0}^{0} \\
\nabla _{0}R^{\alpha}_{\beta} &=& (\ddot{H}+6H\dot{H})\delta^{\alpha}_{%
\beta}+\dddot{C}^{\alpha}_{\beta}+3H\ddot{C}^{\alpha}_{\beta}+3\dot{H}\dot{C}%
^{\alpha}_{\beta} - \frac{\Delta}{a^2}\dot{C}^{\alpha}_{\beta} +2H \frac{%
\Delta}{a^2}C^{\alpha}_{\beta} \\
\nabla _{\beta}R^{0}_{\alpha} &=& a^2 \left[-2H\dot{H}\delta_{\alpha
\beta}+H \ddot{C}_{\alpha \beta}+(3H^2-2\dot{H})\dot{C}_{\alpha \beta}
-H\left(4\dot{H}+\frac{\Delta}{a^2}\right)C_{\alpha \beta} \right] \\
\nabla _{\beta}R^{\beta}_{0} &=& 6H\dot{H} \\
\nabla _{c}R^{c}_{0} &=& 3(\ddot{H}+4H\dot{H}) \\
\nabla _{\beta}R^{\beta}_{\alpha} &=& 0 \\
\Box R^{0}_{0}&=&-3(\dddot{H}+5H\ddot{H}+2\dot{H}^2+2H^2\dot{H}) \\
\Box R^{0}_{\alpha} &=& 0 \, = \, \Box R^{\alpha}_{0} \\
\Box R^{\alpha}_{\beta} &=& -(\dddot{H}+9H\ddot{H}+6\dot{H}^2+22H^2\dot{H}%
)\delta^{\alpha}_{\beta}  -\left(\ddot{D}^{\alpha}_{\beta}+3H\dot{D}^{\alpha}_{\beta}-\left(2H^2+%
\frac{\Delta}{a^2}\right)D^{\alpha}_{\beta}\right) -8H\dot{H}\dot{C}%
^{\alpha}_{\beta} \\
D^{\alpha}_{\beta} & \equiv &  \ddot{C}^{\alpha}_{\beta}+3H\dot{C}^{\alpha}_{\beta}-\frac{\Delta}{a^2}C^{\alpha}_{\beta} \\
\Box R &=&-6\left(\dddot{H}+7H\ddot{H}+4\dot{H}^2+12H^2\dot{H}\right)\end{eqnarray*}
and for any scalar function $f(t)$ of time only, it holds that
\begin{eqnarray*}
f(t)_{;}\,^{\alpha}\!_{\beta} &=& -(H \delta^{\alpha}\!_{\beta} +\dot{C}%
^{\alpha}\!_{\beta})\dot{f} \\
f(t)_{;}\,^{0}\!_{0} &=& - \ddot{f} \\
f(t)_{;}\,^{\alpha}\!_{0} &=& 0 \,= \, f(t)_{;}\,^{0}\!_{\alpha} \\
\Box f(t) &=&-\ddot{f} -3H\dot{f}
\end{eqnarray*}
Thus, since $Y=Y(t)$,
\begin{eqnarray*}
\Box(Y^{n-1}) &=& (1-n)Y^{n-3}\left((n-2)\dot{Y}^2 +Y\ddot{Y}+3HY\dot{Y}%
\right) .
\end{eqnarray*}

\section{Relevant vector quantities using t as the time variable} \label{app:v}
Using $t$ as
the time variable, $H=\frac{\dot{a}}{a}$ and $B^{(\alpha}\,\!_{,\beta )} \equiv
\frac{1}{2}\left(B^{\alpha}\,\!_{,\beta }+ B_{\beta}\,\!^{|\alpha }\right)$, the important vector quantities to linear order in the perturbation are:

\begin{eqnarray*}
\Gamma^{0}_{00}&=&0 \\
\Gamma^{0}_{0\alpha} &=&-aHB_\alpha \\
\Gamma^{0}_{\alpha \beta}&=& a^2H\delta_{\alpha \beta}+aB_{(\alpha ,\beta)}
\\
\Gamma^{\alpha}_{00}&=&-\frac{1}{a}(HB^\alpha+\dot{B}^\alpha) \\
\Gamma^{\alpha}_{0 \beta}&=& H \delta^{\alpha}_{\beta} -\frac{1}{2a}\left(
B^{\alpha}\,\!_{,\beta }-B_{\beta}\,\!^{|\alpha }\right) \\
\Gamma^{\alpha}_{\beta \gamma}&=& aH B^\alpha \delta_{\beta \gamma} \\
\Gamma^{c}_{\alpha c} &=& 0 \\
R_0^0 &=& 3(\dot{H} +H^2) \\
R^0_{\alpha} &=& -\frac{\Delta B_\alpha}{2a} \\
R_0^{\alpha} &=& \left(2\dot{H}+\frac{\Delta}{2a^2}\right)\frac{B^\alpha}{a}
\\
R^{\alpha}_{\beta} &=& (\dot{H} +3H^2)\delta^{\alpha}_{\beta} +\frac{1}{a^3}%
\left(a^2 B^{(\alpha}\,\!_{,\beta )} \right)^{.} \\
R &=& 6(\dot{H}+2H^2) \\
Y\equiv R^a_bR^b_a &=& 12(\dot{H}^2 +3\dot{H}H^2 +3H^4) \\
R^{cd}R^{0}\!_{cd0} &=& -3(\dot{H}+H^2)(\dot{H}+3H^2) \\
R^{cd}R^{0}\!_{cd\alpha} & =& H^2\frac{\Delta}{a} B_\alpha \\
R^{cd}R^{\alpha}\!_{cd\beta} &=& -(3\dot{H}^2+8\dot{H}H^2+9H^4)\delta^{%
\alpha}_{\beta}-\frac{1}{a}(3\dot{H}+2H^2)\dot{B}^{(\alpha}\,\!_{,\beta )} -%
\frac{4}{a}(H\dot{H}+H^3) B^{(\alpha}\,\!_{,\beta )}
\end{eqnarray*}
\begin{eqnarray*}
\nabla _{0}R^{0}_{0} &=& 3(\ddot{H} +2H\dot{H}) \\
\nabla _{0}R^{0}_{\alpha} &=& 2aH\dot{H}B_\alpha -\frac{\Delta}{2a}\dot{B}%
_\alpha+H\frac{\Delta}{a}B_\alpha \\
\nabla _{0}R^{\alpha}_{0} &=& O(B) \\
\nabla _{0}R^{\alpha}_{\beta} &=& (\ddot{H}+6H\dot{H})\delta^{\alpha}_{%
\beta}+\frac{1}{a^3}\left(a^2B^{(\alpha}\,\!_{,\beta )}\right)^{..}-\frac{3H%
}{a^3}\left(a^2B^{(\alpha}\,\!_{,\beta )}\right)^{.} \\
\nabla _{\beta}R^{0}_{0} &=& aH\left(2\dot{H}+\frac{\Delta}{a^2}%
\right)B_\beta \\
\nabla _{\beta}R^{0}_{\alpha} &=& -2a^2H\dot{H}\delta_{\alpha \beta}+a\left[H%
\dot{B}_{(\alpha,\beta )}-2(\dot{H}-H^2)B_{(\alpha ,\beta)} -\frac{\Delta}{%
2a^2}B_{\alpha ,\beta} \right] \\
\nabla _{\beta}R^{\beta}_{0} &=& 6H\dot{H} \\
\nabla _{c}R^{c}_{0} &=& 3(\ddot{H}+4H\dot{H}) \\
\nabla_{\alpha}R^{\beta}_{\gamma} &=& \frac{1}{a^3}\left(a^2B^{(\beta}\,%
\!_{,\gamma ) ,\alpha}\right)^{.} -aH\left(2\dot{H}B^\beta \delta_{\alpha
\gamma}+\frac{\Delta}{2a^2}\left(B^{\beta} \delta_{\alpha \gamma}+B_{\gamma}
\delta^{\beta}_{\alpha}\right)\right) \\
\nabla _{\beta}R^{\beta}_{\alpha} &=& \frac{\Delta}{2a}\dot{B}_{\alpha}
-aH\left(2\dot{H}+\frac{\Delta}{a^2}\right)B_{\alpha} \\
\nabla _{c}R^{c}_{\alpha} &=& 0 \\
\Box R^{0}_{0}&=&-3(\dddot{H}+5H\ddot{H}+2\dot{H}^2+2H^2\dot{H}) \\
\Box R^{0}_{\alpha} &=& \frac{\Delta}{2a}\ddot{B}_\alpha+H\frac{\Delta}{2a}%
\dot{B}_\alpha-2(\dot{H}+H^2)\frac{\Delta}{a}B_\alpha-\frac{\Delta^2}{2a^3}%
B_\alpha \\
\Box R^{\alpha}_{\beta} &=& -(\dddot{H}+9H\ddot{H}+6\dot{H}^2+22H^2\dot{H}%
)\delta^{\alpha}_{\beta}-\frac{1}{a^3}\left(a^2B^{(\alpha}\,\!_{,\beta
)}\right)^{...}+\frac{3H}{a^3}\left(B^{(\alpha}\,\!_{,\beta )}\right)^{..}
\notag \\
&& +\frac{1}{a^3}\left(3\dot{H}+2H^2+\frac{\Delta}{a^2}\right)\left(a^2B^{(%
\alpha}\,\!_{,\beta )}\right)^{.}-\frac{1}{a}\left(8H\dot{H}+2H\frac{\Delta}{%
a^2}\right)B^{(\alpha}\,\!_{,\beta )} \end{eqnarray*}
and for any scalar function $f(t)$ of time only,
\begin{eqnarray*}
f(t)_{;}\,^{0}\!_{0} &=& - \ddot{f} \\
f(t)_{;}\,^{0}\!_{\alpha} &=& 0 \\
f(t)_{;}\,^{\alpha}\!_{0} &=& -\frac{1}{a}B^\alpha \left(\ddot{f}-H\dot{f}%
\right) \\
f(t)_{;}\,^{\alpha}\!_{\beta} &=& -\left(H \delta^{\alpha}\!_{\beta} +\frac{1%
}{a}B^{(\alpha}\!\,_{,\beta)}\right) \dot{f} \\
\Box f(t) &=&-\ddot{f} -3H\dot{f}
\end{eqnarray*}
Note that $R=R(t)$ and $Y=Y(t)$ are the same as in the gravitational wave
case.

\newpage

\section{Relevant scalar perturbation quantities} \label{app:s}

Taking $t$ as the time variable, the metric takes the form
\begin{equation*}
ds^2 = -(1+2\alpha)dt^2-2a\beta_{,\alpha}dt dx^{\alpha}
+a^2(t)\left(\delta_{\alpha \beta}(1+2\phi)+2\gamma_{,\alpha
\beta}\right)dx^{\alpha}dx^{\beta}
\end{equation*}
The important scalar quantities to linear order in the perturbation (using $H=\frac{%
\dot{a}}{a}$, $\chi \equiv a(\beta+a \dot{\gamma})$) are:
\begin{eqnarray*}
\Gamma^{0}_{00}&=&\dot{\alpha} \\
\Gamma^{0}_{0\alpha} &=&(\alpha-aH\beta)_{,\alpha} \\
\Gamma^{0}_{\alpha \beta}&=& a^2\left[\delta_{\alpha
\beta}\left(H-2H\alpha+2H\phi+\dot{\phi}\right)+\left(2H\gamma +\frac{\chi}{%
a^2}\right)_{,\alpha \beta}\right] \\
\Gamma^{\alpha}_{00}&=&\frac{1}{a^2}(\alpha - aH\beta-a\dot{\beta})^{|\alpha}
\\
\Gamma^{\alpha}_{0 \beta}&=& H \delta^{\alpha}_{\beta}+ \dot{\phi}%
\delta^{\alpha}_{\beta} +\dot{\gamma}^{|\alpha}\,\!_{\beta} \\
\Gamma^{\alpha}_{\beta \gamma}&=& aH \beta^{|\alpha} \delta_{\beta \gamma}
+\phi_{, \gamma} \delta^{\alpha}_{\beta}+\phi_{, \beta}\delta^{\alpha}_{
\gamma} - \phi^{|\alpha}\delta_{\beta \gamma} +\gamma_{,\beta
\gamma}\,\!^{\alpha} \\
\Gamma^{c}_{0 c} &=& 3H+\dot{\alpha}+3\dot{\phi}+\Delta\dot{\gamma} \\
\Gamma^{c}_{\alpha c} &=& (\alpha+3\phi+\Delta \gamma)_{,\alpha} \\
g^{cd}\Gamma^{0}_{cd} &=& 3H-\dot{\alpha}-6H\alpha +3\dot{\phi}+\frac{\Delta%
}{a^2}\chi \\
g^{cd}\Gamma^{\alpha}_{cd} &=& \frac{1}{a^2}(-\alpha+a\dot{\beta}%
+2aH\beta-\phi+\Delta \gamma)^{|\alpha} \\
R_0^0 &=& 3(\dot{H} +H^2) -3H\dot{\alpha} -6(\dot{H}+H^2)\alpha - \frac{%
\Delta}{a^2}\alpha +3 \ddot{\phi} +6H\dot{\phi} +\frac{\Delta}{a^2}\dot{\chi}
\\
R^0_{\alpha} &=& -2(H\alpha-\dot{\phi})_{,\alpha} \\
R^{\alpha}_{0} &=& \frac{2}{a^2}(H\alpha -\dot{\phi} +a\dot{H}%
\beta)^{|\alpha} \\
R^{\alpha}_{\beta} &=& \left(\dot{H} +3H^2-H\dot{\alpha}-2(\dot{H}%
+3H^2)\alpha+\ddot{\phi}+6H\dot{\phi}-\frac{\Delta}{a^2}\phi +H\frac{\Delta}{%
a^2}\chi\right)\delta^{\alpha}_{\beta}  \notag \\
&& +\frac{1}{a^2}(-\alpha-\phi+\dot{\chi}+H\chi)^{|\alpha}\!\,_{\beta} \\
R &=& 6(\dot{H}+2H^2) -6H\dot{\alpha}-12(\dot{H}+2H^2)\alpha -2 \frac{\Delta%
}{a^2}\alpha+6\ddot{\phi}+24H\dot{\phi}-4\frac{\Delta}{a^2}\phi  \notag \\
&& +2\frac{\Delta}{a^2}\dot{\chi}+4H\frac{\Delta}{a^2}\chi \\
Y\equiv R^a_bR^b_a &=& 12(\dot{H}^2 +3H^2\dot{H} +3H^4) -12H(2\dot{H}+3H^2)%
\dot{\alpha}-48(\dot{H}^2+3H^2\dot{H}+3H^4) \alpha  \notag \\
&& -4(2\dot{H}+3H^2)\frac{\Delta}{a^2}\alpha +12(2\dot{H}+3H^2)\ddot{\phi}%
+72H(\dot{H}+2H^2)\dot{\phi} -8(\dot{H}+3H^2)\frac{\Delta}{a^2}\phi  \notag
\\
&& +4(2\dot{H}+3H^2)\frac{\Delta}{a^2}\dot{\chi}+8H(\dot{H}+3H^2)\frac{\Delta%
}{a^2}\chi \\
& \equiv & \overline{Y} +\delta Y \, \, \, \text{with} \\
\overline{Y} & \equiv & 12(\dot{H}^2 +3H^2\dot{H} +3H^4) \\
R^{cd}R^{0}\!_{cd0} &=& -3(\dot{H}^2+4H^2\dot{H}+3H^4) +6H(\dot{H}+2H^2)\dot{%
\alpha}+12(\dot{H}^2+4H^2\dot{H}+3H^4)\alpha  \notag \\
&&+ 2(\dot{H}+2H^2)\frac{\Delta}{a^2}\alpha -6(\dot{H}+2H^2)\ddot{\phi}-12H(2%
\dot{H}+3H^2)\dot{\phi}+4(\dot{H}+H^2)\frac{\Delta}{a^2}\phi  \notag \\
&&-2(\dot{H}+2H^2)\frac{\Delta}{a^2}\dot{\chi} -4H(\dot{H}+H^2)\frac{\Delta}{%
a^2}\chi \\
R^{cd}R^{0}\!_{cd\alpha} & =& 4H^2(H\alpha - \dot{\phi})_{,\alpha}
\end{eqnarray*}
\begin{eqnarray*}
R^{cd}R^{\alpha}\!_{cd\beta} &=& -(3\dot{H}^2+8\dot{H}H^2+9H^4)\delta^{%
\alpha}_{\beta} +\delta^{\alpha}_{\beta}\biggl[(6H\dot{H}+8H^3)\dot{\alpha}%
+(12\dot{H}^2+32H^2\dot{H}+36H^4)\alpha  \notag \\
&& +(\dot{H}+2H^2)\frac{\Delta}{a^2}\alpha -(6\dot{H}+8H^2)\ddot{\phi}-(16H%
\dot{H}+36H^3)\dot{\phi} +(\dot{H}+6H^2)\frac{\Delta}{a^2}\phi  \notag \\
&& -(\dot{H}+2H^2)\frac{\Delta}{a^2}\dot{\chi} -(H\dot{H}+6H^3)\frac{\Delta}{%
a^2}\chi \biggr]  \notag \\
&& +\frac{1}{a^2}\biggl[(3\dot{H}+2H^2)\alpha +(\dot{H}+2H^2)\phi-(3\dot{H}%
+2H^2)\dot{\chi}-H(\dot{H}+2H^2)\chi \biggr]^{|\alpha}\,\!_{\beta} \\
\nabla _{0}R^{0}_{0} &=& 3(\ddot{H} +2H\dot{H}) -3H\ddot{\alpha}-(9\dot{H}%
+6H^2)\dot{\alpha} - 6(\ddot{H}+2H\dot{H})\alpha -\frac{\Delta}{a^2}\dot{%
\alpha} +2H\frac{\Delta}{a^2}\alpha  \notag \\
&& +3 \dddot{\phi} +6H\ddot{\phi}+6\dot{H}{\phi}+\frac{\Delta}{a^2}\ddot{\chi%
} -2H\frac{\Delta}{a^2}\dot{\chi} \\
\nabla _{0}R^{0}_{\alpha} &=& \left[ -2H\dot{\alpha}+(2H^2-4\dot{H})\alpha+2%
\ddot{\phi}-2H\dot{\phi}+2aH\dot{H}\beta \right]_{,\alpha} \\
\nabla _{0}R^{\alpha}_{0} &=& \frac{1}{a^2}\left[ 2H\dot{\alpha}+(4\dot{H}%
-2H^2)\alpha -2\ddot{\phi}+2H\dot{\phi}+2a(\ddot{H}-H\dot{H})\beta \right]%
^{|\alpha} \\
\nabla _{0}R^{\alpha}_{\beta} &=& (\ddot{H}+6H\dot{H})\delta^{\alpha}_{%
\beta}+\delta^{\alpha}_{\beta}\biggl[-H\ddot{\alpha}-3(\dot{H}+2H^2)\dot{%
\alpha}-2(\ddot{H}+6H\dot{H})\alpha  \notag \\
&& +\dddot{\phi}+6H\ddot{\phi}+6\dot{H}\dot{\phi} - \frac{\Delta}{a^2}\dot{%
\phi}+2H \frac{\Delta}{a^2}\phi+H \frac{\Delta}{a^2}\dot{\chi} +(\dot{H}%
-2H^2) \frac{\Delta}{a^2}\chi \biggr]  \notag \\
&&+ \frac{1}{a^2}\left[-\dot{\alpha}+2H\alpha-\dot{\phi}+2H\phi +\ddot{\chi}%
-H\dot{\chi}+(\dot{H}-2H^2)\chi \right]^{| \alpha}\,\!_{\beta} \\
\nabla _{\alpha}R^{0}_{0} &=& \left[-3H\dot{\alpha}-(6\dot{H}+2H^2)\alpha-
\frac{\Delta}{a^2}\alpha +3\ddot{\phi}+2H\dot{\phi} + \frac{\Delta}{a^2}\dot{%
\chi} +2aH\dot{H}\beta \right]_{,\alpha} \\
\nabla _{\alpha}R^{0}_{\beta} &=& a^2 \delta_{\alpha \beta}\biggl[-2H\dot{H}%
+2H^2 \dot{\alpha} +8H\dot{H}\alpha +H \frac{\Delta}{a^2}\alpha-2H\ddot{\phi}%
-2\dot{H}\dot{\phi}-4H\dot{H}\phi -H \frac{\Delta}{a^2}\phi  \notag \\
&& -H \frac{\Delta}{a^2}\dot{\chi}+H^2 \frac{\Delta}{a^2}\chi \biggr]+\left[%
-3H\alpha +2\dot{\phi}-H\phi+H\dot{\chi}+(H^2-2\dot{H}) \chi-4a^2H\dot{H}%
\gamma \right]_{, \alpha \beta} \\
\nabla_{\alpha}R_{0}^{\beta} &=&\delta_{\alpha}^{\beta}\left[ 2H\dot{H}-2H^2%
\dot{\alpha}-4H\dot{H}\alpha-H \frac{\Delta}{a^2}\alpha +2H\ddot{\phi}+2\dot{%
H}\dot{\phi}+H \frac{\Delta}{a^2}\phi+H \frac{\Delta}{a^2}\dot{\chi}-H^2
\frac{\Delta}{a^2}\chi\right]  \notag \\
&&+ \frac{1}{a^2}\left[3H\alpha-2\dot{\phi}+H\phi-H\dot{\chi}+(2\dot{H}%
-H^2)\chi\right]^{|\beta}\,\!_{\alpha} \\
\nabla _{c}R^{c}_{0} &=& 3(\ddot{H}+4H\dot{H}) -3H\ddot{\alpha}-(9\dot{H}%
+12H^2)\dot{\alpha}-6(\ddot{H}+4H\dot{H})\alpha- \frac{\Delta}{a^2}\dot{%
\alpha}+2H \frac{\Delta}{a^2}\alpha  \notag \\
&& +3\dddot{\phi}+12H\ddot{\phi}+12\dot{H}\dot{\phi}-2 \frac{\Delta}{a^2}%
\dot{\phi}+4H \frac{\Delta}{a^2}\phi+ \frac{\Delta}{a^2}\ddot{\chi}+(2\dot{H}%
-4H^2) \frac{\Delta}{a^2}\chi \\
\nabla_{\gamma}R_{\beta}^{\alpha} &=&\delta^{\alpha}_{\beta}\left[-H\dot{%
\alpha}-2(\dot{H}+3H^2)\alpha +\ddot{\phi}+6H\dot{\phi}- \frac{\Delta}{a^2}%
\phi+H \frac{\Delta}{a^2}\chi \right]_{, \gamma}  \notag \\
&& -2H\delta_{\beta \gamma}(H\alpha-\dot{\phi}+a\dot{H}\beta)^{|\alpha}
-2H\delta^{\alpha}_{\gamma}(H\alpha-\dot{\phi})_{,\beta}+ \frac{1}{a^2}%
(-\alpha-\phi+\dot{\chi}+H\chi)^{|\alpha}\,\!_{\beta \gamma} \\
\nabla _{c}R^{c}_{\alpha} &=& \left[-3H\dot{\alpha}-6(\dot{H}+2H^2)\alpha-%
\frac{\Delta}{a^2}\alpha+3\ddot{\phi}+12H\dot{\phi}-2\frac{\Delta}{a^2}\phi+%
\frac{\Delta}{a^2}\dot{\chi}+2H\frac{\Delta}{a^2}\chi \right]_{, \alpha}
\end{eqnarray*}
\begin{eqnarray*}
\Box R^{0}_{0}&=&-3(\dddot{H}+5H\ddot{H}+2\dot{H}^2+2H^2\dot{H}) +3H\dddot{%
\alpha}+(12\dot{H}+15H^2)\ddot{\alpha}+(18\ddot{H}+57H\dot{H}+6H^3)\dot{%
\alpha}  \notag \\
&&+12(\dddot{H}+5H\ddot{H}+2\dot{H}^2 +2H^2\dot{H})\alpha +\frac{\Delta}{a^2}%
\ddot{\alpha}-4H\frac{\Delta}{a^2}\dot{\alpha} -(8\dot{H}+4H^2)\frac{\Delta}{%
a^2}\alpha-\frac{\Delta^2}{a^4}\alpha -3\ddddot{\phi}  \notag \\
&&-15H\dddot{\phi}-6(2\dot{H}+H^2)\ddot{\phi}-3(5\ddot{H}+4H\dot{H})\dot{\phi%
}+3\frac{\Delta}{a^2}\ddot{\phi}-2H\frac{\Delta}{a^2}\dot{\phi} +8H^2\frac{%
\Delta}{a^2}\phi-\frac{\Delta}{a^2}\dddot{\chi}  \notag \\
&&+H\frac{\Delta}{a^2}\ddot{\chi}+2(\dot{H}+3H^2)\frac{\Delta}{a^2}\dot{\chi}
-(3\ddot{H}-2H\dot{H}+8H^3)\frac{\Delta}{a^2}\chi +\frac{\Delta^2}{a^4}\dot{%
\chi} \\
\Box R^{0}_{\alpha} &=& \Biggl[2H\ddot{\alpha}+6(\dot{H}+H^2)\dot{\alpha}+(6%
\ddot{H}+12H\dot{H}-16H^3)\alpha-2H\frac{\Delta}{a^2}\alpha  \notag \\
&& -2\dddot{\phi}-6H\ddot{\phi}+16H^2\dot{\phi}+2\frac{\Delta}{a^2}\dot{\phi}%
-4H\frac{\Delta}{a^2}\phi+(4H^2-2\dot{H})\frac{\Delta}{a^2}\chi \Biggr]%
_{,\alpha} \\
\Box R^{\alpha}_{\beta} &=& -(\dddot{H}+9H\ddot{H}+6\dot{H}^2+22H^2\dot{H}%
)\delta^{\alpha}_{\beta}+ \delta^{\alpha}_{\beta} \Biggl[H\dddot{\alpha} +(4%
\dot{H}+9H^2)\ddot{\alpha}+(6\ddot{H}+39H\dot{H}+22H^3)\dot{\alpha}  \notag
\\
&&+(4\dddot{H} +36H\ddot{H}+24\dot{H}^2+88H^2\dot{H})\alpha-H\frac{\Delta}{%
a^2}\dot{\alpha}-(2\dot{H}+4H^2)\frac{\Delta}{a^2}\alpha -\ddddot{\phi}-9H%
\dddot{\phi}  \notag \\
&&-(12\dot{H}+22H^2)\ddot{\phi} -(9\ddot{H}+44H\dot{H})\dot{\phi}+2\frac{%
\Delta}{a^2}\ddot{\phi}+5H\frac{\Delta}{a^2}\dot{\phi}-(2\dot{H}+4H^2)\frac{%
\Delta}{a^2}\phi-\frac{\Delta^2}{a^4}\phi  \notag \\
&&-H\frac{\Delta}{a^2}\ddot{\chi}-(2\dot{H}+H^2)\frac{\Delta}{a^2}\dot{\chi}%
-(2\ddot{H}+3H\dot{H}-4H^3)\frac{\Delta}{a^2}\chi+H\frac{\Delta^2}{a^4}\chi %
\Biggr]  \notag \\
&&+\frac{1}{a^2}\Biggl[\ddot{\alpha}-H\dot{\alpha}-(2\dot{H}+12H^2)\alpha -%
\frac{\Delta}{a^2}\alpha +\ddot{\phi}+7H\dot{\phi} -(2\dot{H}+4H^2)\phi -%
\frac{\Delta}{a^2}\phi  \notag \\
&&-\dddot{\chi}+5H^2\dot{\chi} -(\ddot{H}+5H\dot{H}-4H^3)\chi+\frac{\Delta}{%
a^2}\dot{\chi}+H\frac{\Delta}{a^2}\chi\Biggr]^{|\alpha} \! \, _{\beta} \\
\Box R &=&-6 \dddot{H}-42H\ddot{H}-24\dot{H}^2-72H^2\dot{H}+6H\dddot{\alpha}%
+(24\dot{H}+42H^2)\ddot{\alpha}+(36\ddot{H}+174H\dot{H}+72H^3)\dot{\alpha}
\notag \\
&& +24(\dddot{H}+7H\ddot{H}+4\dot{H}^2+12H^2\dot{H})\alpha+2\frac{\Delta}{a^2%
}\ddot{\alpha} -8H\frac{\Delta}{a^2}\dot{\alpha}-(16\dot{H}+28H^2)\frac{%
\Delta}{a^2}\alpha -2\frac{\Delta^2}{a^4}\alpha-6\ddddot{\phi}  \notag \\
&&-42H\dddot{\phi} -24(2\dot{H}+3H^2)\ddot{\phi}-(42\ddot{H}+144H\dot{H})%
\dot{\phi}+10\frac{\Delta}{a^2}\ddot{\phi} +20H\frac{\Delta}{a^2}\dot{\phi}%
-8(\dot{H}+H^2)\frac{\Delta}{a^2}\phi-4\frac{\Delta^2}{a^4}\phi  \notag \\
&& -2\frac{\Delta}{a^2}\dddot{\chi}-2H\frac{\Delta}{a^2}\ddot{\chi}-(4\dot{H}%
-8H^2)\frac{\Delta}{a^2}\dot{\chi}-(10\ddot{H}+12H\dot{H}-8H^3)\frac{\Delta}{%
a^2}\chi+2\frac{\Delta^2}{a^4}\dot{\chi}+4H\frac{\Delta^2}{a^4}\chi
\end{eqnarray*}
and, for a general scalar function $f(\mathbf{x},t)$,
\begin{eqnarray*}
f_{;00} &=& \ddot{f}- \dot{\alpha}\dot{f} \\
f_{;0 \alpha} &=&\dot{f}_{,\alpha}-(\alpha-aH\beta)_{,\alpha}\dot{f}-Hf_{,
\alpha} \\
f_{;\alpha \beta} &=& f_{,\alpha \beta}- a^2\left[\delta_{\alpha
\beta}\left(H-2H\alpha+2H\phi+\dot{\phi}\right)+\left(2H\gamma +\frac{\chi}{%
a^2}\right)_{,\alpha \beta}\right]\dot{f} \\
f^{; 0}\!\,_{0}&=& -\ddot{f}+2\alpha \ddot{f}+\dot{\alpha}\dot{f} \\
f^{; 0}\!\,_{\alpha}&=& -\dot{f}_{,\alpha}+\alpha_{,\alpha}\dot{f}%
+Hf_{,\alpha} \\
f^{; \alpha}\!\,_{0}&=& \frac{1}{a^2}\left[\dot{f}^{|\alpha}-\alpha^{|\alpha}%
\dot{f}+aH\beta^{|\alpha}\dot{f}-a\beta^{|\alpha}\ddot{f}-Hf^{|\alpha}\right]
\\
f^{; \alpha}\!\,_{\beta}&=& \frac{1}{a^2}f^{| \alpha}\!\,_{\beta} -\left[%
\delta^{\alpha}_{\beta}\left(H-2H\alpha+\dot{\phi}\right) + \frac{1}{a^2}%
\chi^{|\alpha}\!\,_{\beta}\right]\dot{f} \\
\Box f &=& - \ddot{f}+2\alpha \ddot{f}+ \frac{\Delta}{a^2}f-(3H-\dot{\alpha}%
-6H\alpha+3\dot{\phi}+\frac{\Delta}{a^2} \chi)\dot{f} .
\end{eqnarray*}

\bigskip


\begin{thebibliography}{99}

\bibitem{barcot} J.D. Barrow and S. Cotsakis, Phys. Lett. B \textbf{214},
515 (1988); K-I. Maeda, Phys. Rev. D \textbf{39}, 3159 (1989).


\bibitem{mid} J.D. Barrow and J. Middleton, Phys. Rev. D \textbf{75}, 123515
(2007).

\bibitem{special} V. M\"{u}ller and H.-J. Schmidt, Gen. Rel. Grav. \textbf{17%
}, 769 (1985); V. M\"{u}ller, Ann. der Physik \textbf{43}, 67 (1986).


\bibitem{bher1} J.D. Barrow and S. Hervik, Phys. Rev. D \textbf{7}4, 124017
(2006).

\bibitem{cotIX} S. Cotsakis, J. Demaret, Y. De Rop and L. Querella, Phys.
Rev. D \textbf{48}, 4595 (1993).

\bibitem{misner} C.W. Misner, Phys. Rev. Lett. \textbf{22}, 1071 (1969).

\bibitem{bkl} V.A. Belinskii, I. Khalatnikov and E.M. Lifshitz, Adv. Phys.
\textbf{19}, 525 (1970).

\bibitem{misnercc} C.W. Misner, Nature \textbf{214}, 40 (1967);  Astrophys. J. \textbf{151}, 431 (1968).

\bibitem{stab} C.B. Collins and J.M. Stewart, Mon. Not. R. astron. Soc.
\textbf{153}, 419 (1971); C.B. Collins and S.W. Hawking, Astrophys. J.
\textbf{181}, 317 (1972); J.D. Barrow and D.H. Sonoda, Phys. Rep. \textbf{139%
}, 1 (1986); J.D. Barrow, Quart. J. R. Astron. Soc. \textbf{23}, 344 (1982).

\bibitem{pen2} R. Penrose, in \textit{General Relativity: an Einstein Survey}%
, eds. S.W. Hawking and W. Israel (Cambridge University Press, Cambridge,
1979); \textit{The Road to Reality}, (Vintage, London, 2005);
J.D. Barrow and S. Hervik, Class Quantum Grav. \textbf{19}, 5173 (2002);
K.P. Tod, Class Quantum Grav. \textbf{20}, 521 (2003); F.C. Mena and P. Tod,
gr-qc/0702057.

\bibitem{bher2} J.D. Barrow and S. Hervik, Phys. Rev. D \textbf{73}, 023007
(2006).

\bibitem{Clifton2} T. Clifton and J.D. Barrow, Phys. Rev. D \textbf{72},
123003 (2005)

\bibitem{Noh:gw} H. Noh and J. Hwang, Phys. Rev. D \textbf{55}, 5222 (1997).

\bibitem{Noh:vort} J. Hwang and H. Noh, Phys. Rev. D \textbf{57}, 2617
(1988).

\bibitem{Noh:scalar} H. Noh and J. Hwang, Phys. Rev. D \textbf{59}, 04750
(1999).

\bibitem{ellis} G.F.R. Ellis, In \textit{General Relativity and Cosmology},
Varenna Lectures in Physics, Ed. R.K. Sachs, Academic Press, New York, (1971)

\bibitem{Clifton3} T. Clifton and J.D. Barrow, Phys. Rev. D \textbf{72},
103005 (2005).
\end{thebibliography}
\end{document}